\documentclass[12pt]{article}

\usepackage{graphicx}

\def\br{\begin{eqnarray}}
\def\er{\end{eqnarray}}
\def\be{\begin{equation}}
\def\ee{\end{equation}}
\def\({\left(}
\def\){\right)}
\def\pa{\partial}
\def\rlx{\relax\leavevmode}
\def\IR{\rlx\hbox{\rm I\kern-.18em R}}
\def\vp{\varphi}
\def\ve{\varepsilon}

\newcommand{\sbr}[2]{\left\lbrack\,{#1}\, ,\,{#2}\,\right\rbrack}

\def\IZ{\rlx\hbox{\sf Z\kern-.4em Z}}
\def\IR{\rlx\hbox{\rm I\kern-.18em R}}
\def\IC{\rlx\hbox{\,$\inbar\kern-.3em{\rm C}$}}
\def\one{\hbox{{1}\kern-.25em\hbox{l}}}

\def\JHEP#1#2#3{{\sl JHEP} {\bf#1} (#2) #3}

\def\PRD#1#2#3{{\sl Phys. Rev.} {\bf D#1} (#2) #3}

\def\AoM#1#2#3{{\sl Ann. of Math.} {\bf #1} (#2) #3}

\def\InvM#1#2#3{{\sl Invent. Math.} {\bf #1} (#2) #3}

\def\IJMPA#1#2#3{{\sl Int. J. Mod. Phys.} {\bf A#1} (#2) #3}

\def\JPAMT#1#2#3{{\sl J. Phys. A: Math. Theor.} {\bf A#1} (#2) #3}

\def\MPLA#1#2#3{{\sl Mod. Phys. Lett.} {\bf A#1} (#2) #3}

\def\Nonl#1#2#3{{\sl Nonlinearity} {\bf #1} (#2) #3}
\def\JCP#1#2#3{{\sl Journal of Computational Physics} {\bf #1} (#2) #3}

\begin{document}

\begin{titlepage}
\vspace*{-1cm}


\vspace{.2in}
\begin{center}
{\large\bf Quasi-integrability in deformed sine-Gordon models and infinite towers of conserved charges}
\end{center}

\vspace{.3cm}

\begin{center}
Harold Blas$^{a}$,  Hector Flores Callisaya $^{b}$

\vspace{.3 in}
\small

\par \vskip .2in \noindent
$^{(a)}$Instituto de F\'\i sica  ;\\
Universidade Federal de  Mato Grosso  \\ 
Bairro Boa Esperan\c ca, Cep 78060-900, Cuiab\'a - MT - Brazil\\

\par \vskip .2in \noindent
$^{(b)}$~Departamento de Matem\'atica;\\
Universidade Federal  de Mato Grosso  \\ 
Bairro Boa Esperan\c ca, Cep 78060-900, Cuiab\'a - MT - Brazil\\

\normalsize
\end{center}

\vspace{.3in}

\begin{abstract}

We have studied the space-reflection symmetries of some soliton solutions of deformed sine-Gordon models in the context of the quasi-integrability concept. Considering a dual pair of anomalous Lax representations of the deformed model we compute analytically and numerically an infinite number of alternating conserved and asymptotically conserved charges through a modification of the usual techniques of integrable field theories. The charges associated to two-solitons with a definite parity under space-reflection symmetry, i.e. kink-kink (odd parity) and  kink-antikink (even parity) scatterings with equal and opposite velocities, split into two infinite towers of conserved and asymptotically conserved charges. For two-solitons without definite parity under space-reflection symmetry (kink-kink and kink-antikink scatterings with unequal and opposite velocities) our numerical results
show the existence of the asymptotically conserved charges only. However, we show that in the center-of-mass reference frame of the two solitons the parity symmetries and their associated set of exactly conserved charges can be restored. Moreover, the positive parity breather-like (kink-antikink bound state) solution exhibits a tower of exactly conserved charges and a subset of charges which are periodic in time. We back up our results with extensive numerical simulations which also demonstrate the existence  of long lived breather-like states in these models. The time evolution has been simulated by the 4th order Runge-Kutta method supplied with non-reflecting boundary conditions. 
\end{abstract} 
\end{titlepage}

\section{Introduction}
\label{sec:intro}
\setcounter{equation}{0}
Solitons and integrable systems play an important role in the study of non-linear phenomena because often they appear in the description of some physical systems. The soliton properties are intimately related to the integrability of the relevant mathematical models in which they arise \cite{babelonbook, faddeev}. Some deformations of these theories have been shown to possess solitary waves that behave in a scattering process in a similar way to true solitons \cite{jhep1, jhep2, jhep3, jhep4, jhep5, jhep7, jhep6, arxiv1}.
The deformations of the relativistic integrable $SU(N)$ Toda \cite{jhep6} (the $N=2$ case is the sine-Gordon (SG) model in disguise \cite{jhep1,jhep3, arxiv1}) and Bullough-Dodd (BD) \cite{jhep4} models have been shown to posses an infinite number of quantities which are not exactly time-independent but are, however,
asymptotically conserved. Similar phenomena have been observed in the deformations of the non-relativistic focusing and defocusing non-linear Schr\"odinger (NLS) model possessing bright and dark solitons \cite{jhep2, jhep5, jhep7}. For earlier observations on related phenomena, such as the elastic scattering of solitons in some non-integrable theories, see e.g.  \cite{hietarinta}. 

Some calculations in solitary wave collisions  have been done in the regimes which
are close to either integrable or related to a particular relationship between the parameters of
the colliding solitons, e.g. high relative velocity, one of the solitons is significantly larger than
the other one, fast thin solitons and slow broad solitons are among the cases considered in the
literature (see e.g. \cite{shift} and references therein). An analytical approach on inelastic solitary wave interactions for a quartic gKdV equation has been considered  showing the absence of a pure 2-soliton solution in a special regime \cite{martel}.

Recently, certain modified defocusing and focusing NLS models \cite{jhep5, jhep7}, with dark and bright soliton solutions, respectively,  have been shown to exhibit the new feature of an infinite tower of exactly conserved charges. For the special case of
two dark (bright) soliton solutions, where the field components are eigenstates of a space-reflection
symmetry, they exhibit an alternating sequence of exactly conserved and asymptotically conserved charges for 
the scattering process of the solitons.  These were the distinguishing new features associated to the non-relativistic deformed defocusing (focusing)
NLS with dark (bright) soliton solutions, as compared to the previous relativistic quasi-integrable models \cite{jhep1, jhep2, jhep3, jhep4, jhep6, arxiv1}.  
 
This work is a continuation of \cite{jhep1}, in which a deformed sine-Gordon model has been used to introduce the quasi-integrability concept. The main result of our paper is that half of the infinite set of quasi-conserved quantities of the deformed sine-Gordon model of ref. \cite{jhep1} are in fact exactly
conserved, provided that some two-soliton configurations are eigenstates (even or odd) of the space-reflection operator.  By considering linear combinations of the asymptotically conserved charges of \cite{jhep1} we show, through analytical and numerical methods, that one tower of them becomes {\sl exactly} conserved and the other
one remains  quasi-conserved after the combination; in this way, we have strengthened the arguments of \cite{jhep1} for two-soliton configurations with definite parity.  Notice that we have dealt with deformations of the sine-Gordon integrable relativistic field theory and reproduced analogous results to the ones in the non-relativistic non-linear Schr\"odinger model, as presented in the recent contributions for defocusing NLS with dark solitons \cite{jhep5} and the focusing NLS with bright solitons \cite{jhep7}.  In addition, in these last references the authors have found,  through numerical simulations,  that the first non-trivial anomaly vanishes even in the cases where the space-time parity and space-parity arguments do not indicate they should vanish. We believe that  the results in \cite{jhep5, jhep7} and the ones in the present paper open the way for new investigations on the nature of the quasi-integrability phenomena.

We consider the both anomalous Lax representations of the deformed sine-Gordon model, and show that the composition of  the space-reflection parity with a special internal symmetry turns out to be a symmetry relating the both anomalous Lax representations. Our analytical and numerical results indicate that the charges associated to two-solitons with a definite parity under space-reflection symmetry, i.e. kink-kink (odd parity) and  kink-antikink (even parity) scatterings with equal and opposite velocities, split into two infinite towers of conserved and asymptotically conserved charges. In the  case of the positive parity breather-like (kink-antikink bound state) solution one has a tower of exactly conserved charges and a sequence of  charges which oscillate around a fixed value. We also show, through numerical simulations, the existence  of long lived breather-like  states in these models \cite{arxiv1}, which in our formulation exhibit a subset of exactly conserved charges.  

However, it seems to be that such parity property is a necessary condition in order to have the sequence of the exactly conserved
charges in the kink-kink and  kink-antikink systems. In fact, as we will show by numerical simulations, there are some  soliton-like configurations without this symmetry in laboratory coordinates (two-solitons without definite parity under space-reflection symmetry: kink-kink and  kink-antikink scatterings with unequal and opposite velocities) which exhibit asymptotically  conserved charges only. However, we show that in the center-of-mass reference frame of the two solitons the parity symmetries are restored, and then their associated set of exactly conserved charges would be constructed. 

In addition, to simulate the time dependence
of field configurations for computing soliton collisions we used the 4th order Runge-Kutta method supplied with  non-reflecting boundary conditions  suitable to allow  the radiation generated as outgoing waves cross the boundary points $x= \pm L$ freely  \cite{nonreflec}. 
Our simulations show that some radiation is produced by the soliton systems and the rate of loss of the energy depends on the initial conditions of the system.    
 
The paper is organized as follows: in section \ref{qzc0} we discuss the dual set of quasi-zero curvature representations introduced in \cite{jhep1}, based on the $sl(2)$
loop algebra, for a real scalar field
theory subjected to a generic potential, and the dual sets of infinite
number of quasi-conservation laws. We discuss the relationship between the space-time
parity and asymptotically conserved charges. In section \ref{sec:symcharges} we introduce the space-reflection parity symmetry and an order two automorphism of the $sl(2)$ loop algebra relating the both dual sets of  quasi-conserved quantities.  We discuss the relationships between the space-reflection parity and the exactly conserved charges. A tower of new exactly conserved charges is constructed for each field configuration possessing a definite space-reflection parity. In section \ref{sec:expansion}
we perform the expansion of the theory (\ref{eq1}) around the
sine-Gordon model in power series on the deformation parameter $\epsilon$, and discuss the interplay between the parity of the solution and its  dynamics.  In section  \ref{kink} we study the space-time and space-reflection symmetries of the kink-antikink, kink-kink and breather solitons of the standard sine-Gordon model. In section \ref{lorentz} the Lorentz transformation is considered in order to study the Lorentz boost transformation of the anomalies and charges. It is shown the vanishing of the anomalies associated to solitary waves. In section \ref{sec:numerical}  we present the results of our numerical simulations which allowed us to compute and study various properties of the kink-kink, kink-antikink and a system involving  a kink and an antikink bound state (breather). In \ref{conclu} we present some conclusions and discussions. The appendix presents useful $\epsilon$-expansions.

\section{The model}
\label{qzc0}
\setcounter{equation}{0}

We consider Lorentz invariant field theories in
$(1+1)$-dimensions with  equation of
motion
\be
\partial^2 \varphi + \frac{\partial\, V\(\varphi\)}{\partial\, \varphi}=0,
\label{eq1}
\ee
where $\vp$ is a real scalar field $\vp$ and  $V( \varphi )$ is the scalar potential and  the operator $\partial^2$ stands for $\( \frac{\partial^2}{\partial t^2}-\frac{\partial^2}{\partial x^2}\)$. 
Let us  consider a deformed  potential \cite{Bazeia, jhep1}
\br
\label{dpot}
V(\vp, q) = \frac{2}{q^2} \tan^2{\vp} [1- |\sin{\vp}|^q]^2,
\er 
where $q$ is a real parameter such that for $q=2$ the potential reduces to the SG potential, i.e. $V(\vp, 2) = \frac{1}{16} [1- \cos{(4 \vp)}]$. So, we introduce the deformation  parameter $\epsilon$ as $q= 2+ \epsilon$, such that in the limit $\epsilon = 0$ one reproduces the SG model. 

In our numerical simulations we will consider the deformed  potential (\ref{dpot}), however most of the analytical and numerical discussions are valid for any positive parity potential under the space-time and space-reflection symmetries of the field $\vp$ defined below.

Next, we  discuss a pair of dual sets of anomalous zero curvature representations of the equations of motion (\ref{eq1}) in order to  construct the new set of exactly conserved charges. So, in the following subsections, we summarize the relevant steps regarding the construction of the  charges and anomalies in the both dual representations \cite{jhep1}, which will be relevant to our discussions in the section 3. 
 
\subsection{A first set of quasi-conserved charges}
\label{fsc}

Consider the first representation through the Lax potentials  
\br
A_{+}&=& \frac{1}{2}\left[ \(\omega^2 \, V -m\)\, b_{1}
  -i\,\omega\, \frac{\pa\,V}{\pa\,\varphi}\,F_1\right], 
\label{pot11}\\
A_{-}&=& \frac{1}{2}\, b_{-1} - \frac{i}{2}\,
\omega\, \partial_{-}\varphi\, F_0. 
\label{pot1}
\er
The Lax potentials $A_{\pm}$ are $sl(2)$ loop algebra valued functions of two variables $x_{\pm}= \frac{1}{2} (t \pm x)$. We follow the notations of the loop algebra commutation relations provided in the appendix A of \cite{jhep1}.   

The curvature of the connection (\ref{pot11})-(\ref{pot1}) can be written as 
\be
F_{+-}\equiv \partial_{+}A_{-}-\partial_{-}A_{+}+\sbr{A_{+}}{A_{-}}= X
\, F_1 -\frac{i\,\omega}{2}\left[\partial^2 \varphi + \frac{\partial\,
    V}{\partial\, \varphi} \right]\,F_0
\label{zc1}
\ee
with
\be
X = \frac{i\,\omega}{2}\,  \partial_{-}\varphi\,
\left[\frac{\pa^2\,V}{\pa\,\varphi^2}+\omega^2\, V-m\right].
\label{x1}
\ee
When the equation of motion (\ref{eq1}) is satisfied one has that the term in (\ref{zc1}) which lies on the direction of the Lie algebra generator $F_0$ vanishes.  

The construction follows through the so-called abelianization procedure \cite{jhep1, mpla}. So, let us consider the gauge transformation  
\be
A_{\mu}\rightarrow a_{\mu}=g\, A_{\mu}\,g^{-1}-\partial_{\mu}g\,
g^{-1},\,\,\,\mbox{with} \,\,\,\,g={\rm exp}\left[\sum_{n=1}^{\infty} \zeta_n\, F_n\right].
\label{g11}
\ee
Note that  the group element $g$ is an exponentiation of generators lying in
the positive grade subspace generated by the $F_n$'s, with $\zeta_n$ being parameters to be determined. So, when the anomaly does not vanish the new curvature becomes    
\be
F_{+-}\rightarrow
g\,F_{+-}\,g^{-1}=\partial_{+}a_{-}-\partial_{-}a_{+}+\sbr{a_{+}}{a_{-}}=
X \, g\, F_1\,g^{-1},
\label{newc}
\ee
where the equations of motion (\ref{eq1}) has been used.

The connection $a_{-}$  can be  decomposed into its graded components; so one has \cite{jhep1}
\br
a_{-}&=&\frac{1}{2}\,b_{-1}
\label{zz1}\\
&-&\frac{1}{2}\,\zeta_1\,\sbr{b_{-1}}{F_1}- \frac{i}{2}\,
\omega\, \partial_{-}\varphi\, F_0\nonumber\\
&-&\frac{1}{2}\,\zeta_2\,\sbr{b_{-1}}{F_2}
+\frac{1}{4}\,\zeta_1^2\,\sbr{\sbr{b_{-1}}{F_1}}{F_1}
- \frac{i}{2}\,
\omega\, \partial_{-}\varphi\, \zeta_1\,\sbr{F_1}{F_0}
-\partial_{-}\zeta_1\,F_1
\nonumber\\
&\vdots&\nonumber\\
&-&\frac{1}{2}\,\zeta_n\,\sbr{b_{-1}}{F_n}+ \ldots. \nonumber
\er
See appendix B in \cite{jhep1} for the first $\zeta_n$'s obtained by requiring that the component in the direction of $F_{n-1}$ cancels out in $a_{-}$. For example, we can set $\zeta_1= \frac{i}{2}\,
\omega\, \partial_{-}\varphi$, and so on. Therefore, the component $a_{-}$ gets rotated into the abelian sub-algebra generated by the set $\{b_{2n+1}\}$. So, one has
\be
\label{am33}
a_{-}=\frac{1}{2}\,b_{-1}+\sum_{n=0}^{\infty}a_{-}^{(2n+1)}\, b_{2n+1}.
\ee 
Since the anomaly term $X$ in (\ref{zc1}) is non-vanishing it is not possible to transform $a_{+}$ into
the abelian sub-algebra generated by the $b_{2n+1}$. So, $a_{+}$ takes the form
\be
\label{apco}
a_{+}=\sum_{n=0}^{\infty}a_{+}^{(2n+1)}\, b_{2n+1}
+\sum_{n=2}^{\infty} c_{+}^{(n)}\,F_n.
\ee 
The first few terms of the coefficients $a_{\pm}^{(2n+1)}$  and  $c_{+}^{(n)}$ are provided in appendix B of ref. \cite{jhep1}. The process follows by defining the components of $g F_{1} g^{-1}$ in the r.h.s. of eq. (\ref{newc}) as
\br
\label{ff1}
g F_{1} g^{-1} = \sum_{n=0}^{\infty} \gamma^{(2n+1)} b_{2n+1} + \mbox{terms lying on the}\, F_{n}\mbox{'s}.
\er
The terms proportional to the $F_n$'s exactly cancel out and the curvature (\ref{newc}) is left with the terms in the direction of the $b_{2n+1}$ only. So, the transformed curvature (\ref{newc}) provides the equation  
\br
\partial_{+}a_{-}^{(2n+1)}-\partial_{-}a_{+}^{(2n+1)}& = & X \gamma^{(2n+1)} \label{qsc0} \\
&\equiv & \beta^{(2n+1)}\qquad\qquad n=0,1,2,\ldots ;
\label{qsc}
\er
where the quantities $\beta^{(2n+1)}$ are linear in the anomaly $X$ given in
(\ref{x1}). 

The eq. (\ref{qsc}) can be written in the $x$ and $t$ coordinates as  
\be
\frac{d\,Q^{(2n+1)}}{d\,t}=- \,\alpha^{(2n+1)}+
a_{t}^{(2n+1)}\mid_{x=-\infty}^{x=\infty} 
\ee
with
\br
Q^{(2n+1)}\equiv \int_{-\infty}^{\infty}dx\,a_{x}^{(2n+1)},\qquad\qquad\qquad
\alpha^{(2n+1)}\equiv \int_{-\infty}^{\infty}dx\,\beta^{(2n+1)}.
\label{cad1}
\er

Since we are interested in finite energy solutions of the theory
(\ref{eq1}) the field
configurations must satisfy the boundary conditions
\br
\partial_{\mu}\vp \rightarrow 0\; ; \qquad\qquad V\(\vp\)\rightarrow \mbox{\rm
  global minimum}\qquad\qquad {\rm as} \qquad  x\rightarrow \pm \infty.
\label{bc}
\er  
Therefore, for finite energy solutions satisfying the boundary condition (\ref{bc}), one has
\br
\frac{d\,Q^{(1)}}{d\,t}= 0\; , \qquad\qquad\qquad
\frac{d\,Q^{(2n+1)}}{d\,t}=- \,\alpha^{(2n+1)}\qquad
n=1,2,\ldots
\label{qsc1}
\er
For $n=0$ one has $\beta^{(1)}=0$, and $Q^{(1)}$, as discussed in section \ref{sec:symcharges}, is related to the energy and momentum.
The charges $Q^{(2n+1)}, n\geq 1 $ are not conserved due to the non-trivial anomaly $\alpha^{(2n+1)}$.  
 
Next we summarize the main result of \cite{jhep1}. If the field $\vp$ is an eigenstate of the space-time reflection around a given point $(x_\Delta , t_\Delta )$ given by
\br
\label{stsym1}
P:\,\,\,\,\,\,\,\,\,\,\, \vp & \rightarrow & -\vp + \mbox{const.}\\
\label{stsym2}
P : (\widetilde{x}\, ,\, \widetilde{t}) & \rightarrow & (-\widetilde{x}\, ,\, -\widetilde{t}),\,\,\,\,\mbox{with} \,\,\,\,\widetilde{x}=x-x_\Delta\, ,\,\mbox{and},\,\, \widetilde{t}=t-t_{\Delta},
\er  
then, the charges $Q^{(2n+1)}$ in (\ref{qsc1}) satisfy the relationship
\br
\label{mirror1}
Q^{(2n+1)}(t=\widetilde{t}_o + t_{\Delta} ) = Q^{(2n+1)}(t = -\widetilde{t}_o + t_{\Delta} ).
\er
For the two-soliton solutions the charges are asymptotically conserved, i.e. they have the same charges before ($\widetilde{t}_o=-\infty$) and after ($\widetilde{t}_o=+ \infty$) the collision $
Q^{(2n+1)}(+ \infty) = Q^{(2n+1)}(-\infty)$. Whereas, for the breather-like solution, the charges oscillate around a fixed value. 
 
\subsection{A second set of quasi-conserved charges}
\label{ssc}

Next we consider  another quasi-zero curvature representation of the
equation of motion (\ref{eq1}) and  a second set of quasi-conserved charges for the theory. The new Lax potentials are obtained from
(\ref{pot11})-(\ref{pot1}) by interchanging $x_{+}$ with $x_{-}$, and by reverting the signs of
the grades of the relevant generators.  Thus we consider the dual Lax potential \cite{jhep1}
\br
\widetilde{A}_{-}&=& \frac{1}{2}\left[ \(\omega^2\,V-m\)\, b_{-1}
-i\,\omega\,  \frac{\pa\,V}{\pa\,\varphi}\,F_{-1}\right], 
\label{pot22}\\
\widetilde{A}_{+}&=& \frac{1}{2}\, b_{1} - \frac{i}{2}\,
\omega\, \partial_{+}\varphi\, F_0.
\label{pot2}
\er
The curvature of such a connection becomes
\br
\label{cur22}
{\widetilde F}_{+-}\equiv \partial_{+}\widetilde{A}_{-}-\partial_{-} \widetilde{A}_{+}+\sbr{\widetilde{  A}_{+}}{\widetilde{A}_{-}}= \widetilde{X} \, F_{-1}
+\frac{i}{2}\,\omega\,\left[\partial^2\vp+\frac{\partial V}{\partial
    \vp}\right]\, F_0,
\er
with
\be
\widetilde{X} = -\frac{i}{2}\,
\omega\, \partial_{+}\varphi\,\left[\frac{\pa^2\,V}{\pa\,\varphi^2}+\omega^2\,
  V-m\right].
\label{x2}
\ee
The construction of the corresponding charges follows the same procedure as in
section \ref{fsc}. One performs the gauge transformation 
\be
\widetilde{A}_{\mu}\rightarrow \widetilde{a}_{\mu}=\widetilde{g}\, \widetilde{A}_{\mu}\,\widetilde{g}^{-1}-\partial_{\mu}\widetilde{g}\, \widetilde{g}^{-1} ,
\label{gp1}
\ee
with the group element being 
\be
\widetilde{g}={\rm exp}\left[\sum_{n=1}^{\infty} \zeta_{-n}\,
  F_{-n}\right]
\label{gp2}.
\ee
So, one has  
\be
\partial_{+} \widetilde{a}_{-}-\partial_{-} \widetilde{a}_{+}+\sbr{\widetilde{a}_{+}}{\widetilde{a}_{-}}= \widetilde{X} \, \widetilde{g}\,
F_{-1}\,\widetilde{g}^{-1},
\label{nct}
\ee
where   the equation of motion (\ref{eq1}) has been used to cancel the
component of $\widetilde{F}_{+-}$ in the direction of $F_0$. The new connection takes the form
\br
\widetilde{a}_{+}&=&\frac{1}{2}\,b_{1}
+\sum_{n=0}^{\infty}\widetilde{a}_{+}^{(-2n-1)}\, b_{-2n-1},
\label{wap1}\\
\widetilde{a}_{-}&=&\sum_{n=0}^{\infty}\widetilde{a}_{-}^{(-2n-1)}\, b_{-2n-1}
+\sum_{n=2}^{\infty} \widetilde{c}_{+}^{(-n)}\,F_{-n}.
\label{wan1}
\er 
We refer to the appendix C of ref. \cite{jhep1} for the first few terms of the coefficients ${\widetilde a}_{\pm}^{(-2n-1)}$ and $\widetilde{c}_{+}^{(n)}$. Let us define the components of $\widetilde{g} F_{-1} \widetilde{g}^{-1}$ in the r.h.s. of eq. (\ref{nct}) as
\br
\label{f221}
\widetilde{g} F_{-1} \widetilde{g}^{-1} = \sum_{n=0}^{\infty} \widetilde{\gamma}^{(-2n-1)} b_{-2n-1} + \mbox{terms lying on the}\, F_{-n}\mbox{'s}.
\er
A similar construction as in the section (\ref{fsc}) allows us to write the curvature (\ref{nct}) as 
\br
\label{qsct1}
\partial_{+}{\widetilde a}_{-}^{(-2n-1)}-\partial_{-}{\widetilde a}_{+}^{(-2n-1)} &=& \widetilde{X}\, \widetilde{\gamma}^{(-2n-1)} \\
& \equiv & {\widetilde \beta}^{(-2n-1)}\qquad\qquad n=0,1,2,\ldots
\label{qsct}
\er
with the quantity ${\widetilde \beta}^{(2n+1)}$ being linear in the anomaly ${\widetilde X}$, given in
(\ref{x2}).

Applying the same boundary conditions as in section \ref{fsc}, for finite energy solutions the quasi conservation laws become
\be
\frac{d\,\widetilde{Q}^{(-1)}}{d\,t}= 0\; , \qquad\qquad\qquad
\frac{d\,\widetilde{Q}^{(-2n-1)}}{d\,t}=- \,{\widetilde
  \alpha}^{(-2n-1)}\qquad 
n=1,2,\ldots
\label{qsc2}
\ee
with
\br
\widetilde{Q}^{(-2n-1)}\equiv \int_{-\infty}^{\infty}dx\,{\widetilde
  a}_{x}^{(-2n-1)},\qquad\qquad\qquad 
\widetilde{\alpha}^{(-2n-1)}\equiv \int_{-\infty}^{\infty}dx\,\widetilde{\beta}^{(-2n-1)}. 
\label{cad2}
\er

Following similar arguments as in the last subsection, one can conclude  that the above construction also admit a set of charges which obey a mirror time-symmetry (\ref{mirror1}) and, consequently, a set of asymptotically conserved charges $
\widetilde{Q}^{(-2n-1)}(+ \infty) = \widetilde{Q}^{(-2n-1)}(-\infty). $   
 
In the next section we will define a new set of infinitely many quasi-conserved charges and relevant anomalies associated to the pair of charges $\(Q^{(2n+1)} ,\, \widetilde{Q}^{(-2n-1)}\)$ and anomalies $\(\alpha^{(2n+1)},\,\widetilde{\alpha}^{(-2n-1)} \)$ by means of their linear combinations. Similar construction has been performed in the usual sine-Gordon model  \cite{chodos} in order to obtain an infinite number of conservation laws written in the space-time coordinates $\(x,t\)$.  The new representations of the charges will turn out to be convenient ones in order to analyse their properties in relation to the space-reflection symmetries of the soliton field configurations.

\section{Space-reflection parity and conserved charges}
\label{sec:symcharges}

We will consider the above dual sets of anomalous Lax representations of the deformed  sine-Gordon model and their associated quasi-conservation laws in order to construct a sequence of conserved charges and vanishing anomalies. The construction below follows by relating the dual sets through an operator which is a composition  of the space-reflection  and an order two automorphism of the $sl(2)$ affine loop algebra. The space-reflection symmetry of some soliton solutions of the deformed SG model will imply the existence of an infinite tower of conserved charges. In order to see this  it is convenient to examine a linear combination, at each order $n=1,2,...$, of the above two sets of quasi-conserved charges $Q^{(2n+1)}$ (\ref{cad1}) and $\widetilde{Q}^{(-2n-1)}$  (\ref{cad2}). The discussion for the particular case $n=0$ will be presented below.  Notice that  each set of the quasi-conserved charges was constructed considering a particular quasi-zero curvature representation of the same deformed sine-Gordon model. So, let us consider the new quasi-conservation laws 
\br
\label{lcom}
\frac{dQ_{\pm}^{(2n+1)}}{dt} = - \alpha_{\pm}^{(2n+1)}, \,\,\,\,n=1,2,...,
\er
with the charges $Q_{\pm}^{(2n+1)}$ and anomalies $\alpha_{\pm}^{(2n+1)}$ being defined as the following linear combinations
\br
\label{charlc}
Q_{\pm}^{(2n+1)} & \equiv &  \mp \frac{1}{w^2} \(Q^{(2n+1)} \pm \widetilde{Q}^{(-2n-1)}\),\\
  \alpha_{\pm}^{(2n+1)} &\equiv & \mp \frac{1}{w^2}  \(\alpha^{(2n+1)} \pm  \widetilde{\alpha}^{(2n+1)}  \)\label{anolc0}\\
 &=&  \mp \frac{1}{w^2} \int_{-\infty}^{+ \infty} dx\, \( \beta^{(2n+1)} \pm  \widetilde{\beta}^{(-2n-1)}\),\label{anolc}
\er
in which the quantities  $Q^{(2n+1)}$   and $ \alpha^{(2n+1)}$  defined  in eqs. (\ref{cad1}) and (\ref{qsc1}) and  the quantities $\widetilde{Q}^{(-2n-1)}$ and  $\widetilde{\alpha}^{(2n+1)} $ in eqs. (\ref{qsc2})-(\ref{cad2})  have been used.

Since  the theory (\ref{eq1}) is invariant under space-time translations one has that the energy
momentum tensor is conserved. This conservation law is related to the vanishing of the anomalies $\alpha_{\pm}^{(1)}=0$ (i.e. the case $n=0$), which follow from the vanishing of the integrand functions $\beta^{(1)}$  and $\widetilde{\beta}^{(-1)}$. In fact, the linear  combinations of the charges $Q^{(1)}$   and $\widetilde{Q}^{(-1)}$, lead to the 
energy and momentum of the field configuration, respectively, i.e.
\br
\label{ener}
Q_{+}^{(1)} &=& \int_{-\infty}^{+ \infty} dx\,\,\Big[\frac{1}{2}(\pa_t \vp)^2 + \frac{1}{2}(\pa_x \vp)^2 + V \Big],  \\
Q_{-}^{(1)} &=& \int_{-\infty}^{+ \infty} dx\,\, \pa_x \vp \pa_t \vp, 
\label{mom}\er
where $E=Q_{+}^{(1)} $ is the energy and $P=Q_{-}^{(1)}$ is the momentum.
 
The first non-trivial anomalies from (\ref{anolc})  become
\br
 \alpha_{\pm}^{(3)} &=& \pm \frac{1}{2}\int_{-\infty}^{+ \infty} dx\, [V'' + w^2 V -m ] \{ \pa_{-} [ (\pa_{-} \vp)^2] \mp \pa_{+} [ (\pa_{+} \vp)^2] \}\\ 
 \alpha_{\pm}^{(5)} &=&  \pm \frac{1}{2}\int_{-\infty}^{+ \infty} dx \,[V'' + w^2 V -m ] \times\\
                    &&  \Big[\(\frac{3}{2} w^2 (\pa_{-}\vp)^2 \pa_{-}^2\vp + \pa_{-}^4\vp\)\pa_{-}\vp \pm \(\frac{3}{2} w^2 (\pa_{+}\vp)^2 \pa_{+}^2\vp + \pa_{+}^4\vp\)\pa_{+}\vp  \Big].
\er
 
The behaviour of the charges $Q_{\pm}^{(2n+1)}$ and anomalies $\alpha_{\pm}^{(2n+1)}$ in the quasi-conservation laws (\ref{lcom}) for soliton collisions will depend on the symmetry properties of the associated field configurations, in particular on the space-reflection symmetry of each anomaly density $ ( \beta^{(2n+1)} \pm  \widetilde{\beta}^{(-2n-1)} )$ of the corresponding anomaly $\alpha_{\pm}^{(2n+1)}$ in (\ref{anolc}), as we will see below. So, let us examine the space-reflection symmetry of those densities. 

Consider the space-reflection transformation 
\br
\label{px}
{\cal P}_x: x_{+} \leftrightarrow x_{-},
\er
and assume that the scalar field is an eigenstate of the operator ${\cal P}_x$ 
\br
\label{pxvp}
{\cal P}_x: \vp \rightarrow \varrho\,\, \vp,\,\,\,\,\,\varrho = \pm 1.
\er
In addition, consider an even potential $V$ under ${\cal P}_x$
\br 
\label{evenpot}
{\cal P}_x (V) = V.
\er

So, the Lax potentials $A_{\pm}$ in (\ref{pot11})-(\ref{pot1}) and $\widetilde{A}_{\pm}$ in (\ref{pot22})-(\ref{pot2}) are related by
\br
\widetilde{\Omega}(A_{\pm}) = \widetilde{A}_{\mp},  \\
  \widetilde{\Omega} \equiv \widetilde{\Sigma} {\cal P}_x,
\er
 where $\widetilde{\Sigma}$ is an order two automorphism of the $sl(2)$   loop algebra 
\br
\nonumber
\widetilde{\Sigma}(  
b_{2n+1}) &=& b_{-2n-1} \\
\widetilde{\Sigma}(  
F_{2n+1}) &=& \varrho\, F_{-2n-1}\label{aut22}\\
\widetilde{\Sigma}(  
F_{2n}) &=& \varrho\, F_{-2n}.\nonumber
\er
Next, let us examine the transformation properties  of the anomalies integrand functions  $\beta^{(2n+1)}$  and $\widetilde{\beta}^{(-2n-1)}$. From (\ref{x1}) and (\ref{x2}) one can see that the functions $X$ and $\widetilde{X}$ are related by the transformation
\br
\label{pxxx}
{\cal P}_x (X) = - \varrho \, \widetilde{X}.
\er
Moreover,  the quantities $\gamma^{(2n+1)}$ (\ref{qsc0}) and $\widetilde{\gamma}^{(-2n-1)}$ (\ref{qsct1}) are related through the transformation
\br
\label{pxg}
{\cal P}_x \( \gamma^{(2n+1)} \) = \varrho  \widetilde{\gamma}^{(-2n-1)}.
\er
The above relationship can be obtained by applying $\widetilde{\Omega}$ on the both sides of the eq. (\ref{ff1}) such that
\br
\label{omf1}
\widetilde{\Omega} \( g F_{1} g^{-1}\) &=& \sum_{n=0}^{\infty} {\cal P}_x \(\gamma^{(2n+1)}\) \widetilde{\Sigma} ( b_{2n+1}) +  \mbox{terms lying on the}\, \varrho F_{-n}\mbox{'s}\\
  \varrho \widetilde{g} F_{-1}   \widetilde{g}^{-1} & = &   \sum_{n=0}^{\infty} {\cal P}_x\(\gamma^{(2n+1)}\) b_{-2n-1} + \mbox{terms lying on the}\, \varrho F_{-n}\mbox{'s} ,   \label{omf2}
\er
and comparing the equation (\ref{omf2}) with  the eq. (\ref{f221}) one gets (\ref{pxg}). In order to  relate the group elements $g$ in (\ref{g11})  and $\widetilde{g}$ in (\ref{gp2}) as  $\widetilde{\Omega} (g) = \widetilde{g}$ we have used the automorphism transformation $\widetilde{\Sigma}$ (\ref{aut22})   of the generators $F_n$ and the transformation  ${\cal P}_x(\zeta_n) =\varrho  \zeta_{-n}$ in (\ref{g11}). This transformation property can also be seen by inspecting the explicit expressions of the first few  $\zeta_{\pm n}'s$ presented in the appendix B and C of  \cite{jhep1}.  

From the relationships (\ref{pxxx}) and (\ref{pxg}) one can get 
\br
{\cal P}_x \( X \gamma^{(2n+1)}\) = -  \widetilde{X} \widetilde{\gamma}^{(-2n-1)} \,\,\, \rightarrow\,\,\, {\cal P}_x \( \beta^{(2n+1)}\) = - \widetilde{\beta}^{(-2n-1)},\,\,\,\,n= 1,2,...
\er
where the eqs. (\ref{qsc0})-(\ref{qsc}) for  $\beta^{(2n+1)}$  and  the eqs. (\ref{qsct1})-(\ref{qsct}) for  $\widetilde{\beta}^{(-2n-1)}$ have been used.
Therefore, for the linear combinations we have
\br
\label{odd1}
{\cal P}_x \(\beta^{(2n+1)} +  \widetilde{\beta}^{(-2n-1)}\) &=& - \(\beta^{(2n+1)} +  \widetilde{\beta}^{(-2n-1)}\), \,\,\,\,n=1,2,...\\
\label{even1}
{\cal P}_x \(\beta^{(2n+1)} -  \widetilde{\beta}^{(-2n-1)}\) &=& + \(\beta^{(2n+1)} -  \widetilde{\beta}^{(-2n-1)}\), \,\,\,\,n=1,2,...
\er 
Therefore, for these type of field configurations satisfying (\ref{pxvp}) we have that the anomaly densities possess a definite space-reflection parity, i.e. the anomaly density of $\alpha_{+}^{(2n+1)}$ ($\alpha_{-}^{(2n+1)}$) is an odd (even) function of $x$. So, from the eq. (\ref{anolc}) and using the symmetry property  (\ref{odd1}) one has that the set of anomalies $\alpha_{+}^{(2n+1)}$ vanish, i.e. 
\br
\label{ano123}
\alpha_{+}^{(2n+1)} = 0, \,\,\,\,n=1,2,...
\er  
This implies that a subset of the quasi-conservation laws (\ref{lcom}) turn out to be truly conservation laws
\br
\label{cons1}
\frac{dQ_{+}^{(2n+1)}}{dt} = 0, \,\,\,\,n=1,2,...,
\er
with the charges $Q_{+}^{(2n+1)}$  defined as 
\br
\label{charge11}
Q_{+}^{(2n+1)} & \equiv &  - \frac{1}{w^2} \(Q^{(2n+1)} + \widetilde{Q}^{(-2n-1)}\), 
\er
in which the charges  $Q^{(2n+1)}$ and  $\widetilde{Q}^{(-2n-1)}$  have been defined in (\ref{cad1}) and  (\ref{cad2}), respectively.

Some comments are in order here: 

First, if the field $\vp$ is an eigenstate (\ref{pxvp}) of the space reflection symmetry (\ref{px}) for a fixed time one can construct a tower of conserved charges $Q_{+}^{(2n+1)}$ in (\ref{cons1}). Since the energy  $E=Q_{+}^{(1)} $ in (\ref{ener}) is a conserved quantity for any solution, it can be included into the sequence of exactly conserved charges (\ref{cons1}), i. e. we will have the infinite sequence of exactly conserved charges   
\br
\label{cons11}
\frac{dQ_{+}^{(2n+1)}}{dt} = 0, \,\,\,\,n=0, 1, 2,...
\er
Second, notice that for an even or odd parity field $\vp$  the momentum  $P=Q_{-}^{(1)}$ in (\ref{mom}) vanishes, since the momentum density is an odd function under (\ref{px}). As we will see below, in their center-of-mass reference frames the kink-kink, kink-antikink and breather (kink-antikink bound state) solutions of the usual sine-Gordon model posses definite parities under (\ref{px}), and so their total momenta will vanish.  

Third, since the charges $Q_{\pm}^{(2n+1)}$ in (\ref{lcom}) are linear combinations of asymptotically conserved charges, provided that the field configurations $\vp$ are eigenstates of the space-time symmetry (\ref{stsym1})-(\ref{stsym2}) as discussed in the last section, we may conclude that the charges $Q_{\pm}^{(2n+1)}$ will also be asymptotically conserved.

Let us write the  integrands of the anomalies $\alpha_{\pm}^{(3)}$ in terms of  the $\pa_x$ and  $\pa_t$ derivatives. So, once the eq. of motion (\ref{eq1}) is used to substitute $\pa_t^2 \vp \rightarrow [\pa_x^2\vp -V'(\vp)]$ in the integrand of $\alpha_{+}^{(3)}$, one has
\br
\label{alf1}
\alpha_{+}^{(3)}=-2 \int \, dx \,[V'' + w^2 V -m ]  \Big\{ \pa_{x} [ (\pa_{t} \vp)^2] + \pa_{x} [ (\pa_{x} \vp)^2] - \pa_x V(\vp)\Big\}.
\er 
Notice that the `surface' terms of the form $\pa_x\{ \star \}$ in the integrand of (\ref{alf1}) can be discarded by taking into account the boundary condition (\ref{bc}). Therefore, one is left with
\br
\label{alf11}
\alpha_{+}^{(3)} & =&-2 \int \, dx \,[V'' + w^2 V]  \Big\{ \pa_{x} [ (\pa_{t} \vp)^2] + \pa_{x} [ (\pa_{x} \vp)^2]\Big\},\\
& \equiv & \int \, dx \, f_{+}^{(3)}(x,t),\label{alf112}
\er
where we have defined the anomaly density $f_{+}^{(3)}$. Notice that for even parity potentials (\ref{evenpot}) and for definite parity (even or odd eigenstate) fields $\vp$ (\ref{pxvp}) the density $f_{+}^{(3)}$ is an odd function, and thus the anomaly $\alpha_{+}^{(3)}$ vanishes. 

Similarly, once the eq. of motion (\ref{eq1}) is used to substitute $\pa_x^2 \vp \rightarrow [\pa_t^2\vp + V'(\vp)]$ in the integrand of $\alpha_{-}^{(3)}$, one has
\br
\label{alf2}
\alpha_{-}^{(3)}=-2 \int \, dx \,[V'' + w^2 V -m ]  \Big\{ \pa_{t} [ (\pa_{t} \vp)^2] + \pa_{t} [ (\pa_{x} \vp)^2] + \pa_t V(\vp)\Big\}.
\er 
Notice that the terms which can be written in the form $\pa_t [  \int \, dx \, ( \star ) ]$ in (\ref{alf2}) can be carried to the l.h.s. of the quasi-conservation law (\ref{lcom}) such that the charge $Q^{(3)}_{-}$ gets redefined as $Q^{(3)}_{-} \rightarrow Q^{(3)}_{-} + \int \, dx \, ( \star )$. Therefore, discarding the terms $[V'' + w^2 V -m ] \pa_t V(\vp)$ and $-m \{\pa_{t}[ (\pa_{t} \vp)^2] + \pa_{t} [ (\pa_{x} \vp)^2] \}$ in the anomaly density above, one is left with the redefined anomaly
\br
\label{alf22}
\alpha_{-}^{(3)} = -2 \int \, dx \,[V'' + w^2 V]  \Big\{ \pa_{t} [ (\pa_{t} \vp)^2] + \pa_{t} [ (\pa_{x} \vp)^2]\Big\}.
\er
Furthermore, once we substitute $\pa_t^2 \vp \rightarrow [\pa_x^2\vp - V'(\vp)]$ into this expression and follow similar reasoning as above it acquires the following form  
\br
\label{alf223}
\alpha_{-}^{(3)} &=& -4 \int \, dx \,[V'' + w^2 V]  \Big\{ \pa_{t} \vp \pa_{x}^2 \vp + \pa_{x} \vp \pa_{x} \pa_t \vp \Big\},\\
&\equiv& \int \, dx \, f_{-}^{(3)}(x,t),\label{alf2233}
\er
where we have defined the anomaly density $f_{-}^{(3)}$. Notice that for even parity potentials (\ref{evenpot}) and for definite parity (even or odd eigenstate) fields $\vp$ (\ref{pxvp}) the density $f_{-}^{(3)}$ is an even function, and thus the anomaly $\alpha_{-}^{(3)}$ will not vanish solely by a space-reflection parity reason. 

Notice that the redefined effective anomalies provided by the expressions (\ref{alf11}) and (\ref{alf223}) will vanish when evaluated on static solutions; in particular, they will vanish for static kink-type solutions. In sec. \ref{vanano} we will show that the anomalies $\alpha_{\pm}^{(3)}$ vanish for traveling solutions, and in particular for traveling kinks,  in all Lorentz frames.  Moreover, the effective expressions (\ref{alf11}) and (\ref{alf223}), being more amenable to numerical simulations, and the relevant time integrated expressions  $\int dt \, \alpha_{\pm}^{(3)}$,  will be computed numerically for various two-soliton and breather-like configurations below.

\section{Perturbation theory and space-reflection symmetry}
\label{sec:expansion}
\setcounter{equation}{0}

Next let us analyse the solutions which admit the equation (\ref{eq1})  such that  $\vp$ satisfies the space-reflection symmetry (\ref{px})-(\ref{pxvp}). In sec. \ref{spref} we will examine the space-reflection symmetry of the two-soliton (kink-kink, kink-antikink and breather) solutions of the usual SG model. We perform this construction in perturbation theory around solutions of the SG model, so let us expand the solutions of (\ref{eq1}) into power series in $\epsilon$, as   
\be
\vp= \vp_0+\vp_1\,\ve +\vp_2\,\ve^2+\ldots.
\ee
The expansion of the potential (\ref{dpot}) is presented in Appendix \ref{expansion}.

Therefore,  expanding the equation of motion (\ref{eq1}) in powers of $\ve$ we find that the
order zero field $\vp_0$ must satisfy the sine-Gordon equation,
{\it i.e.} 
\be
\partial^2\vp_0+\frac{1}{4}\,\sin\(4\,\vp_0\)=0.
\label{vp0}
\ee
The higher order component of the field $\vp_n$ satisfies the
equation  
\be
\partial^2\vp_n+ \cos{(4 \vp_0)}\, \vp_n = f_n
\label{fn},
\ee   
where the result $\frac{\partial^2 V}{\partial \vp^2}|_{\ve=0} = \cos{(4 \vp_0)}$ for the deformed potential (\ref{dpot}) has been used. The first few $f_n$'s are provided in the appendix \ref{expansion}.
 
Next, we split (\ref{fn}) into the even and odd parts under the space-reflection parity ${\cal P}_x$ such that
\br
\label{fnpm}
\pa^2 \vp_{n}^{(\pm)} + \Big[\cos{( 4\, \vp_0)}\, \vp_n\Big]^{(\pm)} = f_n^{(\pm)},\,\,\,\,\mbox{where}\,\,\,\, \star^{(\pm)} \equiv \frac{1}{2} \(1\pm {\cal P}_x \) \star,\,\,\,\,\,n=1,2,...
\er
In the next section we consider the two sectors separately and label them by the parameter $\varrho$ in (\ref{pxvp}).  It is assumed that $\vp_0$  possesses a definite parity under the transformation (\ref{px}). 

\subsection{The case $\varrho = 1$ }    

In this case one has from  (\ref{pxvp}) that $\vp_0^{+} \neq 0$ and $\vp_0^{-}=0$, and so
\br
\label{order0}
{\cal P}_x : \vp_0 \rightarrow +  \vp_0.    
\er
 Therefore, $\vp_0$ assumes an even parity and (\ref{f11}) implies that $f_1^{(+)} \neq 0$ and $f_1^{(-)}=0$. So, the first order equations become
\br
\label{f1p}
\pa^2 \vp_{1}^{(+)} + \cos{( 4\, \vp_0^{(+)})}\,\, \vp_1^{(+)} &=& f_1^{(+)}(\vp_0^{(+)}),\\
\pa^2 \vp_{1}^{(-)} + \cos{( 4\, \vp_0^{(+)})}\,\, \vp_1^{(-)} &=& 0.\label{f11p}
\er
Some comments are in order here. First, the pair of fields $\vp^{(+)}_1$ and  $\vp_{1}^{(-)}$ satisfies the uncoupled linear system of equations (\ref{f1p})-(\ref{f11p}) with variable coefficients. Second, the equation of motion for the even component $\vp_{1}^{(+)}$ (\ref{f1p})   satisfies a non-homogeneous equation and so it can never vanish.  However, the odd part component $\vp_{1}^{(-)}$ (\ref{f11p}) satisfies a homogeneous equation and so it can vanish. Third, if the field without definite parity $\vp_{1}$ is a solution, so is the field combination $\vp_{1}-\vp_{1}^{(-)}=\vp_{1}^{(+)}$. Then, one can always choose a first order solution which is even under the space-reflection parity, i.e.
\br
\label{order1}
{\cal P}_x : \vp_1 \rightarrow \vp_1.    
\er
Next, choosing the zeroth (\ref{order0}) and first order (\ref{order1}) solutions one has that $f_2$ in (\ref{f22}) splits into $f_2^{(+)} \neq 0$ and
 $f_2^{(-)} = 0$. So, the second order terms  ${\cal O}(\epsilon^2)$ in  (\ref{fnpm}) become
\br
\label{f2p}
\pa^2 \vp_{2}^{(+)} + \cos{( 4\, \vp_0^{(+)})}\,\, \vp_2^{(+)} &=& f_2^{(+)}(\vp_0^{(+)}, \vp_1^{(+)} ),\\
\pa^2 \vp_{2}^{(-)} + \cos{( 4\, \vp_0^{(+)})}\,\, \vp_2^{(-)} &=& 0.\label{f22p}
\er
Similarly, as in the first order case, from (\ref{f2p})-(\ref{f22p}) one can always choose a second order solution which is even under the space-reflection parity, i.e.
\br
\label{order2}
{\cal P}_x : \vp_2 \rightarrow \vp_2.    
\er
Following similar reasoning as above one notices that choosing (\ref{order0}), (\ref{order1}) and  (\ref{order2}) solutions the function $f_3$ in (\ref{f33}) splits into $f_3^{(+)} \neq 0$ and
 $f_3^{(-)} = 0$, and then one can write for the third order solution 
\br
\label{order3}
{\cal P}_x : \vp_3 \rightarrow \vp_3.    
\er  
So, one can choose again the third order solution to be even, and this construction can be  repeated in all orders ${\cal O}(\epsilon^n)$ in order to  construct a perturbative solution which satisfies (\ref{px})-(\ref{pxvp})  with $\varrho=1$ , and so has
charges satisfying (\ref{cons1}). Therefore, the theory (\ref{eq1}) possesses a subset of solutions such that the charges $Q^{(2n+1)}_{+},\,n=1,2,3,...$ are exactly  conserved. It must be included into this series the charges $ Q^{(1)}_{\pm}$ presented in (\ref{ener})-(\ref{mom}).

\begin{figure}
\centering
\label{fig1}
\includegraphics[width=12cm,scale=3, angle=0,height=4.5cm]{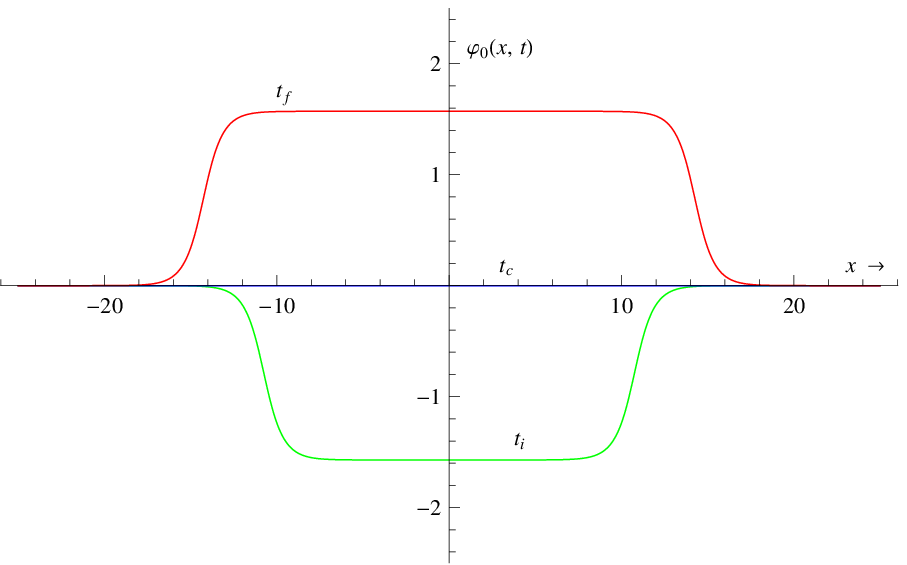}
\includegraphics[width=12cm,scale=3, angle=0,height=4.5cm]{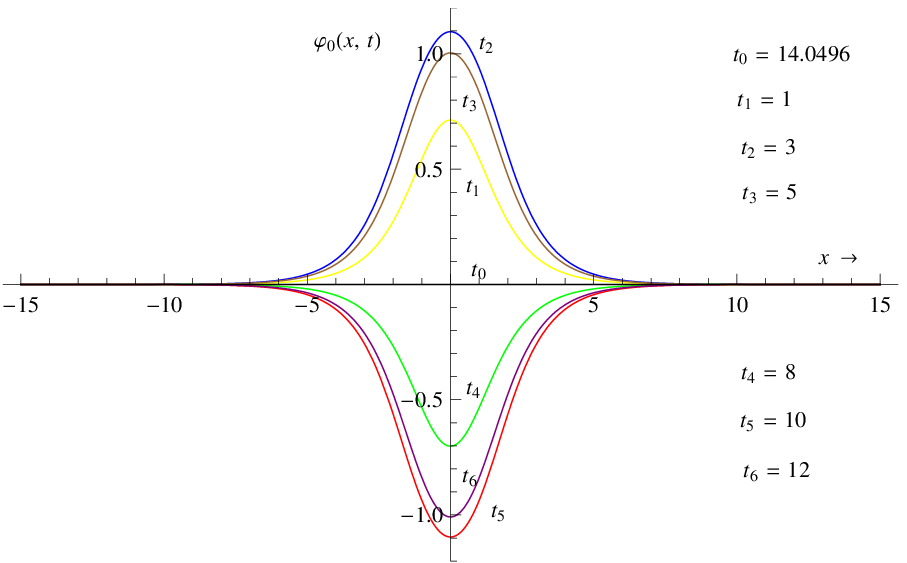}
\parbox{6in}{\caption{(color online) a) $\vp_0$ with even space-reflection parity for an antikink/kink pair sent at $v_1=-v_2=0.7$, for initial $t_i$ (green), collision  $t_c$ (horizontal black line, i.e. $\vp(x, t_c)=0$) and final $t_f$ (red) times, respectively. At $t_c$ the solitons have merged and cancel to each other. After collision, $t_f$, each soliton emerges as kink or antikink, respectively. b)  $\vp_0$ with even space-reflection parity for the SG breather oscillating with period $T=14.0496$ plotted for successive times $t_0< t_2....< t_6$, respectively. At $t_0 = T$ the soliton/antisoliton pair has merged and they cancel to each other.}}
\end{figure}

\subsection{The case $\varrho = -1$ }

In this case one has from  (\ref{pxvp}) that $\vp_0^{+} = 0$ and $\vp_0^{-} \neq 0$, and so
\br
\label{order00}
{\cal P}_x : \vp_0 \rightarrow -\vp_0.    
\er 
Therefore, $\vp_0$ assumes an odd parity and (\ref{f11}) implies that $f_1^{(+)} = 0$ and $f_1^{(-)} \neq 0$. So, the first order equations become
\br
\label{f1m}
\pa^2 \vp_{1}^{(+)} + \cos{( 4\, \vp_0^{(-)})}\,\, \vp_1^{(+)} &=& 0,\\
\pa^2 \vp_{1}^{(-)} + \cos{( 4\, \vp_0^{(-)})}\,\, \vp_1^{(-)} &=& f_1^{(-)}(\vp_0^{(-)}).\label{f11m}.
\er
It follows  similar arguments as above. First, the pair of fields $\vp^{(-)}_1$ and  $\vp_{1}^{(+)}$ satisfies the uncoupled linear system of equations (\ref{f1m})-(\ref{f11m}) with variable coefficients. Second, the equation of motion for the odd component $\vp_{1}^{(-)}$ (\ref{f11m})   satisfies a non-homogeneous equation and so it can never vanish.  However, the even part component $\vp_{1}^{(+)}$ (\ref{f1m}) satisfies a homogeneous equation and so it can vanish. Third, if the field without definite parity $\vp_{1}$ is a solution, so is the field combination $\vp_{1}-\vp_{1}^{(+)}=\vp_{1}^{(-)}$. Then, one can always choose a first order solution which is odd under the space-reflection parity, i.e.
\br
\label{order11}
{\cal P}_x : \vp_1 \rightarrow - \vp_1 .   
\er
Next, choosing the zeroth (\ref{order00}) and first order (\ref{order11}) solutions one has that $f_2$ in (\ref{f22}) splits into $f_2^{(+)} = 0$ and
 $f_2^{(-)} \neq 0$. So, the second order terms ${\cal O}(\epsilon^2)$ in  (\ref{fnpm}) become
\br
\label{f2pp}
\pa^2 \vp_{2}^{(+)} + \cos{( 4\, \vp_0^{(+)})}\,\, \vp_2^{(+)} &=& 0,\\
\pa^2 \vp_{2}^{(-)} + \cos{( 4\, \vp_0^{(+)})}\,\, \vp_2^{(-)} &=& f_2^{(-)}(\vp_0^{(-)}, \vp_1^{(-)} ).\label{f22pp}
\er
Similarly, as in the first order case, from (\ref{f2pp})-(\ref{f22pp}) one can always choose a second order solution which is odd under the space-reflection parity, i.e.
\br
\label{order22}
{\cal P}_x : \vp_2 \rightarrow -\vp_2.    
\er
Next, choosing (\ref{order00}), (\ref{order11}) and  (\ref{order22}) solutions the function $f_3$ in (\ref{f33}) splits into $f_3^{(+)} = 0$ and
 $f_3^{(-)} \neq 0$, and then one can write for the third order solution 
\br
\label{order33}
{\cal P}_x : \vp_3 \rightarrow -\vp_3.    
\er  
One can choose again the third order solution to be odd, and this process can be  repeated order by order with $\vp_n$ satisfying (\ref{px})-(\ref{pxvp}) with $\varrho=-1$, and so one has charges satisfying (\ref{cons1}). Therefore, the theory (\ref{eq1}) possesses a subset of solutions such that the charges $Q^{(2n+1)}_{+},\,n=1,2,3,...$ are exactly  conserved. It must also be included into this series the charges $ Q^{(1)}_{\pm}$ provided in (\ref{ener})-(\ref{mom}).

Let us summarize the main results so far. The main result of \cite{jhep1} is that the deformed SG model presents an infinite number of asymptotically conserved charge, for each anomalous zero-curvature representation, and for solitons satisfying the space-time symmetry (\ref{stsym1})-(\ref{stsym2}). Next, for solitons satisfying the same space-time symmetry, as well as the space-reflection symmetry (\ref{px})-(\ref{pxvp}) one can say more. In this case, the sequence of the charges $\{Q^{(1)}_{\pm},\, Q^{(2n+1)}_{+},\,n=1,2,...\}$ become indeed exactly conserved (\ref{cons1}), and so, it constitutes a new result for deformed models of the SG type. So, the model supports an infinite number of conserved charges $\{Q^{(1)}_{\pm},\, Q^{(2n+1)}_{+},\,n=1,2,...\}$ and asymptotically conserved charges $\{Q^{(2n+1)}_{-},\,n=1,2,...\}$  associated to soliton solutions  satisfying the both space-time and space-reflection symmetries.  

\begin{figure}
\centering
\label{fig2}
\includegraphics[width=12cm,scale=3, angle=0,height=4.5cm]{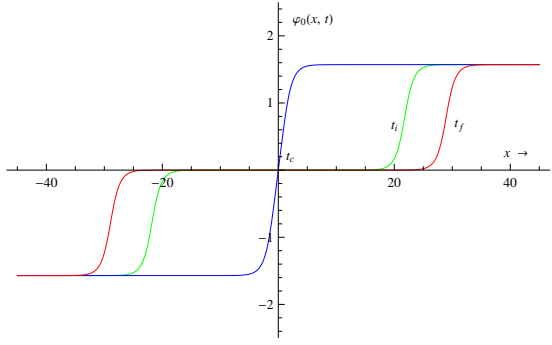}
\parbox{6in}{\caption{(color online) $\vp_0$ with  odd space-reflection parity  for a kink/kink pair sent at $v_1=-v_2=0.7$, for initial $t_i$ (green), collision  $t_c$ (blue) and final $t_f$ (red) times, respectively. After collision, $t_f$, each soliton emerges as a kink.}}
\end{figure}

\begin{figure}
\centering
\label{fig3}
\includegraphics[width=12cm,scale=3, angle=0,height=4.5cm]{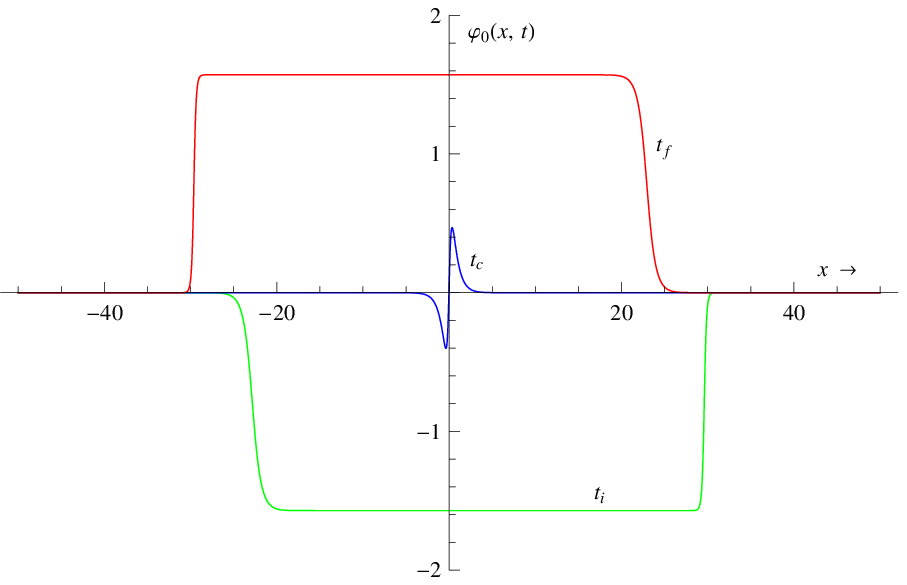}
\includegraphics[width=12cm,scale=3, angle=0,height=4.5cm]{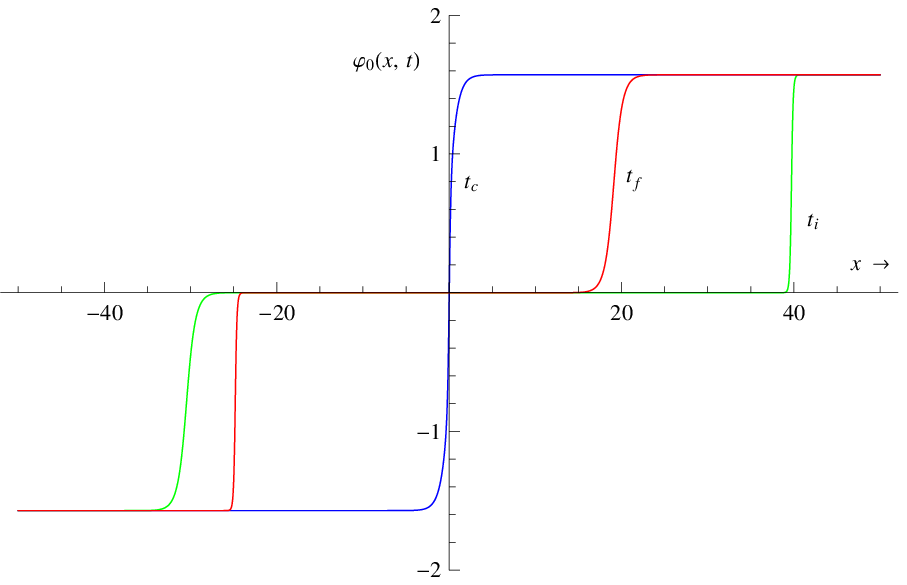}
\parbox{6in}{\caption{(color online) a)  $\vp_0$ without definite space-reflection parity for an antikink/kink pair sent at $v_1=0.017, v_2=-0.044$, for initial $t_i$ (green), collision  $t_c$ (blue) and final $t_f$ (red) times, respectively. After collision, $t_f$, each soliton emerges as kink or antikink, respectively. b) $\vp_0$ without definite space-reflection parity for a kink/kink pair sent at $v_1=0.017, v_2=-0.049$, for initial $t_i$ (green), collision  $t_c$ (blue) and final $t_f$ (red) times, respectively. After collision, $t_f$, each soliton emerges as a kink.}}
\end{figure}

\section{Space-time symmetries of sine-Gordon solitons}

\label{kink}

Next we discuss the both space-time and space-reflection symmetries in the general two-soliton and breather solutions of the integrable sine-Gordon model. The solutions of the SG model (\ref{vp0}) can be found in terms of the tau functions as \cite{babelon, jhep00}
\br
\label{vp0tau}
\vp_0 = - \frac{i}{2} \log{\frac{\tau_0}{\tau_1}}.
\er
Since $\vp_0$ is real we must have $\tau_1 =\tau_0^{\star}$ and so
\br
\label{vp0arctan}
\vp_0 = \arctan{\frac{Im(\tau_0)}{Re(\tau_0)}},
\er
where the $\tau_0$ function is given by\footnote{Since $\vp_0 \rightarrow \vp_0 + \pi/2$ leaves invariant the SG eq. (\ref{vp0}), the relevant expression of \cite{jhep1} can be rewritten as $e^{2 i \vp_0} = \frac{-i\, \tau_0}{i \, \tau_1}$, so the form of the tau function $\tau_0$ in (\ref{tau0}) follows from that given in \cite{jhep1} provided that the transformation $\tau_0 \rightarrow -i \tau_0$, $\tau_1 \rightarrow i \tau_1$ is performed.}
\br
\label{tau0}
\tau_0 = -i + e^{\Gamma_1} + e^{\Gamma_2} + i \gamma e^{\Gamma_1 + \Gamma_2} ,
\er
with
\br
\Gamma_i &=& \epsilon_i \frac{x- v_i t -x_0^{(i)}}{\sqrt{1-v_i ^2}},\,\,\,\,\,i=1,2 \\
\gamma &=& \Big[\tanh{(\frac{\alpha_2-\alpha_1}{2})}\Big]^{2 \epsilon_1 \epsilon_2},
\er
where  $\epsilon_i = \pm 1$, for kink and anti-kink, and $v_i = \tanh{\alpha_i}$, is the velocity (since $c=1$, one must have $v_i < 1 $) of the kink (antikink) $i$.

The solution (\ref{vp0arctan}) can be written as
\br
\label{vp0sch}
\vp_0 = \arctan{\frac{\sqrt{\gamma} \sinh{z_+}}{\cosh{z_{-}}}},
\er 
where
\br
z_{+} = \frac{\Gamma_1 + \Gamma_2}{2}+\log{\sqrt{\gamma}},\,\,\,\,\,z_{-} = \frac{\Gamma_1 - \Gamma_2}{2}.
\er
Then, the space-time parity transformation (\ref{stsym2}) in the new variables $z_{\pm}$ becomes
\br
P\,:\,\,\,\,\, (z_{+}\,,\,z_{-}) \,\,\rightarrow \,\,   (-z_{+} \,,\, -z_{-}) ,
\er
and the solution $\vp_0$ in (\ref{vp0sch}), taking the domain of $\arctan$ to be $(-\pi/2,\pi/2)$, transforms as 
\br
P(\vp_0) = - \vp_0.
\er
This space-time symmetry has been presented in \cite{jhep1}.  

\subsection{Space-reflection symmetry of sine-Gordon solitons}
\label{spref}
Next, let us examine the space-reflection symmetry  (\ref{px})-(\ref{pxvp}) of the two-soliton solution (\ref{vp0sch}). From  (\ref{vp0sch}) we expect to recover the well known kink-kink, kink-antikink and breather type solutions (see e.g. \cite{rajaraman}). So, we will be interested on some set of parameters $\{v_1,v_2,\epsilon_1, \epsilon_2 \}$ such that the  two-soliton solution provides the field $\vp_0$ with a definite parity under the   space-reflection symmetry for any given shifted time $\widetilde{t}=t-t_{\Delta}$, i.e 
\br
\label{pxxy}
{\cal P}_x &:& x \rightarrow -x,\\
\label{pxx1}
{\cal P}_x &: & \vp_0 (x, t)\rightarrow \varrho\,\, \vp_0(x,t),\,\,\,\,\,\varrho = \pm 1.
\er

For $\epsilon_1=-\epsilon_2 = +1,\, v_2=-v_1$ one has the kink-antikink solution
\br
\label{sa1}
\vp_0^{SA} = \arctan{\Big[\frac{1}{v_1} \frac{\sinh{(\frac{v_1 t}{\sqrt{1-v_1^2}})}}{\cosh{(\frac{x}{\sqrt{1-v_1^2}})}}\Big]}.
\er

Notice that the kink-antikink solution possesses an even parity ($\varrho =+1$) under the space-reflection transformation (\ref{pxxy})-(\ref{pxx1}), and it can be regarded as the zeroth order solution of deformed SG model in perturbation theory, as elaborated in sec. \ref{sec:expansion}.  This configuration is plotted in Fig. 1a for three successive times. 

For $\epsilon_1=\epsilon_2 = +1,\, v_2=-v_1$ one has the kink-kink solution
\br
\label{ss1}
\vp_0^{SS} = \arctan{\Big[v_2 \frac{\sinh{(\frac{x}{\sqrt{1-v_2^2}})}}{\cosh{(\frac{v_ 2 t}{\sqrt{1-v_2^2}})}}\Big]}.
\er

Notice that the kink-kink solution possesses an odd parity  ($\varrho =-1$)  under the space-reflection transformation (\ref{pxxy})-(\ref{pxx1}), and constitutes a good candidate to be the zeroth order solution of deformed SG model in perturbation theory, as elaborated in sec. \ref{sec:expansion}. This configuration is plotted in Fig. 2 for three successive times.

Making the transformation $v_1 \rightarrow i \nu/\sqrt{1-\nu^2}$ in (\ref{sa1}), i.e. convert the real parameter $v_1$ into an imaginary one, one has 
\br
\label{br1}
\vp_0^{b} = \arctan{\Big[\frac{\sqrt{1-\nu^2}}{\nu} \frac{\sin{( \nu\, t )}}{\cosh{(\sqrt{1-\nu^2}\, x)}}\Big]},
\er
where $|\nu|< 1$ is its frequency. This is another real solution for $\vp_0$, and it is interpreted as a ``bound" solution of a kink-antikink pair at rest. This solution is the well known breather solution  of the SG model and it is a periodic in time solution with period $T_0 = 2\pi/\nu$.
 
Notice that the breather solution at rest satisfies $\vp_0^{b} (-x, t) = \vp_0^{b}(x, t)$ and therefore it  possesses an even parity  ($\varrho =+1$)  under the space-reflection transformation (\ref{pxxy})-(\ref{pxx1}), and it could also be used as the zeroth order solution of the deformed SG model in perturbation theory as considered in sec. \ref{sec:expansion}. This configuration is plotted in Fig. 1b for various times $0<t<T_0$.

Moreover, it is possible to have two-soliton solutions without definite space-reflection parity
\br
\label{2kasym1}
\vp_0^{SA} &=& \arctan{\Big\{\frac{1}{\tanh{(\frac{\alpha_1-\alpha_2}{2})}} \frac{\sinh{\Big[\frac{1}{2} \( \frac{t\, v_1 - x}{\sqrt{1-v_1^2}}- \frac{t\, v_2 -x}{\sqrt{1-v_2^2}} \)\Big]}}{\cosh{\Big[\frac{1}{2} \( \frac{t\, v_1 - x}{\sqrt{1-v_1^2}}+ \frac{t\, v_2 -x}{\sqrt{1-v_2^2}} \)\Big]}} \Big\}},\,\,\,\,\,\\
\label{2kasym2}
\vp_0^{SS}&=& \arctan{\Big\{\tanh{(\frac{\alpha_2-\alpha_1}{2})} \frac{\sinh{\Big[\frac{1}{2} \( \frac{t\, v_1 - x}{\sqrt{1-v_1^2}}+ \frac{t\, v_2 -x}{\sqrt{1-v_2^2}} \)\Big]}}{\cosh{\Big[\frac{1}{2} \( \frac{t\, v_1 - x}{\sqrt{1-v_1^2}} - \frac{t\, v_2 -x}{\sqrt{1-v_2^2}} \)\Big]}} \Big\}},
\er
where $v_1 = \tanh{\alpha_1},\,v_2 = \tanh{\alpha_2}$. In  Fig. 3  we plot these two-solitons. In Fig. 3a the case $\epsilon_1=-\epsilon_2 = +1,\, v_2 \neq -v_1$ is plotted for three successive times, which is an asymmetric  kink-antikink pair. In Fig 3b the case $\epsilon_1=\epsilon_2 = +1,\, v_2 \neq -v_1$ is plotted as an asymmetric  kink-kink solution for three successive times. However, one can transform the above solutions through the Lorentz boosts $(x,t)\rightarrow (x', t')$ in order to write them in the   forms (\ref{sa1}) and (\ref{ss1}) for the kink-antikink and kink-kink, respectively. The relevant Lorentz boosts $(x,t)\rightarrow (x', t')$ become
\br
\label{lor1}\mbox{kink-antikink}: x' &=& -x\cosh{\chi}  + t \sinh{\chi},\,\, t' = -x \sinh{\chi}  + t \cosh{\chi},\\
 \mbox {kink-kink}: x' &=& -x \cosh{\chi}  + t \sinh{\chi}\, ,\,\, t' = x \sinh{\chi}  - t \cosh{\chi} ,\label{lor2}
\er    
where $\chi \equiv (\alpha_1+\alpha_2)/2$. In  the new coordinates $(x', t')$ the two-solitons will be parity eigenstates under the space-reflection: $x' \rightarrow - x';\, t' \rightarrow  t' $ transformation.

The last explicit examples show that the space-reflection parity symmetries of the kink-antikink and kink-kink  solutions are not, in general, Lorentz invariant, i.e. ${\cal P}_{x'} \neq \Lambda {\cal P}_x  \Lambda^{-1}$ with $\Lambda$  being a Lorentz transformation. In fact, in the above examples we have shown that the kink-kink and kink-antikink solutions are eigenstates of ${\cal P}_x$ only in their centre of mass reference frames, respectively.

\section{Lorentz  transformations of charges and anomalies}
\label{lorentz}

Let us consider the two dimensional Lorentz transformation
\br
\Lambda :   \,\,\,\,x_{\pm} \rightarrow e^{\mp \kappa}  x_{\pm},  
\er
where the rapidity $\kappa$ and the velocity $v$ are related by $ v= \tanh{\kappa}$. And, in addition let $\sigma$ be an automorphism  of the loop algebra $sl_{2}$ 
\br
\sigma(T) = e^{\kappa d}\, T\, e^{-\kappa d},
\er
where $d$ is the grading operator. So, it follows that under the composition of the above transformations, the Lax operators of the both representations (\ref{pot11})-(\ref{pot1}) and (\ref{pot22})-(\ref{pot2}), respectively,  transform as vectors   
\br
\Theta (A_{\pm}) = e^{\pm \kappa} A_{\pm},\,\,\,\,\,\,\, \Theta ({\widetilde A}_{\pm}) = e^{\pm \kappa} {\widetilde A}_{\pm},\,\,\,\,\, \Theta \equiv \Lambda \sigma.
\er 
Therefore, the curvatures (\ref{zc1}) and (\ref{cur22}) are invariant under the composed transformation $\Theta$, and so are the anomalous terms $X F_1$ and $\widetilde{X} F_{-1}$, respectively.  In order to see the Lorentz transformation laws of the charges and anomalies in (\ref{qsc1}) and (\ref{qsc2}) we need to examine the properties of the quantities $\gamma^{(2n+1)}$ introduced in (\ref{ff1}) and $\widetilde{\gamma}^{(-2n-1)}$ in (\ref{f221}), respectively.

Let us examine the relevant quantities of the first gauge transformation performed in section \ref{fsc}.  So, the gauge transformation (\ref{g11}) of the connection  $A_{-}$ in (\ref{pot1}) gives rise to the new connection $a_{-}$, which has been decomposed in (\ref{zz1}). The term $\frac{i}{2} w \pa_{-}\vp F_{0}$ in  the r.h.s. of the second equation of (\ref{zz1}) undergoes a transformation 
\br
  \Theta(\frac{i}{2} w \pa_{-}\vp F_{0}) = e^{-\kappa} \,\, \frac{i}{2} w \pa_{-}\vp F_{0}.
\er

Since we have chosen the parameter $\zeta_1$  such that the both terms in the second eq. of (\ref{zz1}) cancel to each other, it follows that
\br
\Theta(\zeta_1) = \Lambda(\zeta_1) = e^{-\kappa} \zeta_1.
\er
This implies that each one of the last three terms on the r.h.s. of  the third eq. of (\ref{zz1}) gets multiplied by $e^{-\kappa}$ under the action of $\Theta$. Therefore, in order to cancel the $F_1$ component in that eq. one must have $\Theta(\zeta_2) = e^{-2 \kappa} \zeta_2$. Continuing an analogous reasoning, order by order, one can show that
 \br
 \label{lzetan}
 \Theta(\zeta_n) = \Lambda(\zeta_n) = e^{-n \kappa} \zeta_n,
\er which implies that $\Theta(\zeta_n F_{n} ) = \zeta_n F_{n}$. So, the group element $g$ related to the gauge transformation (\ref{g11}) becomes invariant
\br
\label{glt}
\Theta(g) = g.
\er    
Therefore, as the connection $A_{\pm}$, the new connection $a_{\pm}$ transforms as a vector 
\br
\Theta \( a_{\pm}\) = e^{\pm \kappa}  a_{\pm}.
\er
From (\ref{glt}) one notices  that $\Theta(g F_{1} g^{-1} ) = e^{\kappa}\, g F_{1} g^{-1}$, and then each term on the r.h.s. of  (\ref{ff1}) turns out  to be multiplied by $ e^{\kappa}$ when it is applied with $\Theta$. Then, since $\Theta(b_{2n+1})= e^{(2n+1)\kappa} b_{2n+1}$, it follows that  
\br
 \Theta(\gamma^{(2n+1)}) = \Lambda(\gamma^{(2n+1)}) = e^{-2 n \kappa} \gamma^{(2n+1)}.
\er
Next, using the relationship  $\Theta(X) = \Lambda(X) = e^{-\kappa} X$ in the eqs. (\ref{qsc0})-(\ref{qsc}) and the defining equation for $\alpha^{(2n+1)}$ (\ref{cad1}), one can write the transformation rule for the expression $\alpha^{(2n+1)} dt$ as 
\br
\label{tensor1}
\Theta\(\alpha^{(2n+1)} dt\) = e^{(-2n-1) \kappa} \(\alpha^{(2n+1)} dt \). 
\er
Therefore, from the quasi-conservation law  (\ref{qsc1}) one has that $-\alpha^{(2n+1)} dt$, and so $d Q^{(2n+1)}$, is a tensor under the two-dimensional Lorentz group. 

A quite analogous procedure as above can be implemented for the second anomalous Lax representation leading to the following  transformation rule for the expression $\widetilde{\alpha}^{(-2n-1)} dt$
\br
\label{tensor2}
\Theta\(\widetilde{\alpha}^{(-2n-1)} dt\) = e^{(2n+1) \kappa} \(\widetilde{\alpha}^{(-2n-1)} dt\). 
\er

Therefore, from the quasi-conservation law (\ref{qsc2}) one has that $-\widetilde{\alpha}^{(-2n-1)} dt$, and so $d \widetilde{Q}^{(-2n-1)}$, is a tensor under the two-dimensional Lorentz group.

\subsection{Solitary waves and vanishing anomalies}
\label{vanano}

For static solutions, i.e. field configurations which are $x-$dependent only, the charges $Q^{(2n+1)}$ and $\widetilde{Q}^{(2n+1)}$ are obviously $t-$independent, therefore their associated anomalies $\alpha^{(2n+1)} $ and $\widetilde{\alpha}^{(-2n-1)} $ vanish, respectively. In addition, traveling solutions can be obtained through a Lorentz boost transformation from the static ones. Next we will show that the solitary waves, being special traveling waves, provide vanishing anomalies even for non-integrable theories and, consequently, the charges $Q^{(2n+1)}_{\pm}$ in (\ref{lcom}) are exactly conserved for traveling solutions. In order to achieve this, we will show that $\alpha^{(2n+1)}_{\pm} dt $ and $d Q^{(2n+1)}_{\pm}$ vanish for traveling waves in  all $(1+1)-$dimensional Lorentz frames. Therefore, if $d Q^{(2n+1)}_{\pm}=0$ in the rest frame of the static solution, they must vanish in all Lorentz frames.

In fact, the expressions  $\alpha^{(2n+1)} dt$ in (\ref{tensor1}) and $\widetilde{\alpha}^{(-2n-1)} dt$ in (\ref{tensor2}) transform as tensors under the $1+1$ Lorentz group, consequently if they vanish in the rest frame of the solution, they must vanish in all Lorentz reference frames. Therefore, if $d Q^{(2n+1)}$ and $d \widetilde{Q}^{(-2n-1)}$ vanish on the rest frame of the solution, they should vanish in
all Lorentz frames. One then concludes that the charges $Q^{(2n+1)}$ and $\widetilde{Q}^{(2n+1)}$ are exactly conserved for
traveling wave solutions (like one-soliton solutions) of (\ref{eq1}). In fact, such conclusion holds 
for any functional of the scalar field $\vp$ and its derivatives, which is a tensor under the
Lorentz group.  Consequently, from (\ref{lcom}) one has that the composed anomalies also vanish in all Lorentz frames, i.e. $\alpha^{(2n+1)}_{\pm}=0$.   Consequently, if $d Q^{(2n+1)}_{\pm}$ vanish on the rest frame of the solution, it should vanish in
all Lorentz frames. One then concludes that the composed charges $Q^{(2n+1)}_{\pm}$ are exactly conserved for
traveling wave solutions, such as the solitary waves, of the model (\ref{eq1}).

We must emphasize  that this property holds despite the fact that the expressions $\alpha_{\pm}^{(2n+1)} dt $ involving the composed anomalies (\ref{anolc0}) do not transform as tensors of the Lorentz group, as it is clear from the transformations rules (\ref{tensor1}) and (\ref{tensor2}) for each term of their linear combinations, i.e. $\alpha_{\pm}^{(2n+1)} dt = (\alpha^{(2n+1)} dt \pm  \widetilde{\alpha}^{(2n+1)} dt )$ in    (\ref{lcom}). The vanishing of the anomalies $\alpha_{\pm}^{(2n+1)}$ merely reflects the fact that each term of their linear combinations vanish in all Lorentz frames.  

\section{Numerical support}
\label{sec:numerical}
\setcounter{equation}{0}

In order to check our results on the  anomalies $\alpha_{\pm}^{(3)}$ we have performed various numerical 
simulations  of the Bazeia at. al. model, studying kink-antikink, kink-kink and a system involving   
a kink and an antikink bound state (breather). We will study the behaviour of the quasi-conservation laws in (\ref{lcom}) through numerical simulations of soliton collisions.  After integration they can be written as
\br
\label{timeint}
Q^{(3)}_{\pm}(t) - Q^{(3)}_{\pm}(t_0) =- \int^t_{t_0} dt' \alpha^{(3)}_{\pm}(t'),
\er
where $t_0$ is the initial time of the simulation, taking to be zero. We will consider the effective anomaly expressions (the ones with ``surface" terms and total time derivatives discarded) as presented in (\ref{alf11})-(\ref{alf112}) and (\ref{alf223})-(\ref{alf2233}), respectively. We used various grid sizes and  number of points. The two-soliton (kink-antikink and kink-kink) simulations were performed on a lattice of $2000$ lattice points with lattice spacing of $\Delta x =0.03$ (in the interval $[-L,L]=[-30,30]$). The time step for all of our simulations was $\Delta t =0.001$. The breather-like simulations were performed on a lattice of $10000$ lattice points with lattice spacing of $\Delta x =0.01$ (in the interval $[-L,L]=[-50,50]$). The time evolution was simulated by the fourth order Runge - Kuta method provided that the so-called non-reflecting (transparent) boundary conditions (see e.g. \cite{nonreflec} and references therein) are assumed at the both ends of the lattice grids
\br
\label{nonreflec}
\frac{\pa  }{\pa t} \vp (\pm L, t) \pm \frac{\pa  }{\pa x} \vp (\pm L, t) =0.
\er
So,  the total energy is not conserved in the interval $[-L,L]$, but the only energy which flows to the outer regions  $|x| > L$ is the energy of radiation waves. The b.c.'s (\ref{nonreflec}) assure that the radiation generated as outgoing waves cross the boundary points $x= \pm L$ freely (it is expected that the radiation leaves the domain $[-L,L]$ without being reflected back), then the total remaining energy is effectively the energy of the field configurations related  to the interacting solitons.
Our simulations show that some radiation is produced by the soliton systems and the rate of loss of the energy depends on the initial conditions and the  parameter values for each system. The results of our extensive numerical simulations confirm the usefulness  of the transparent boundary condition (\ref{nonreflec}).   
 
\subsection{kink-antikink}
 
The simulations of the kink-antikink system of the deformed SG model will consider, as the initial condition, two analytical solitary wave solutions. In fact, in order to have a  kink-antikink system for $t=0$ we consider a kink ($\eta_1= 1,\eta_2=1, l=0$) and an antikink ($\eta_1=-1,\eta_2=1, l=0$), according to the solution in eq. (1.2) of \cite{jhep1}, located some distance
apart and stitched together  at the middle point $x=0$.  

These kink-antikink simulations  are presented in the Figs. 4-5. In the Fig. 4 we show  the results for the 
collision of equal and opposite velocity  solitons with  parameters $v_2=-v_1=0.5$ and $\epsilon = 0.06$.  The relevant anomaly densities $f^{(3)}_{\pm}(x,t)$, anomalies $\alpha^{(3)}_{\pm}(t)$, as presented in (\ref{alf11})-(\ref{alf112}) and (\ref{alf223})-(\ref{alf2233}), respectively,  and the time integrated anomalies $\int^t dt' \alpha^{(3)}_{\pm}(t')$ as functions of time, are plotted. The relevant anomaly densities $f^{(3)}_{\pm}(x,t)$ in (\ref{alf112}) and (\ref{alf2233}) have been plotted as functions of $x$ for three successive times (top figures of each column). These plots show qualitatively the behaviour of these functions such that their integration in the whole space furnish vanishing  $\alpha^{(3)}_{+}(t)$ and non-vanishing  $\alpha^{(3)}_{-}(t)$ anomalies (middle figures in the both columns). The bottom figures of the left columns show the vanishing expressions $\int^t dt' \alpha^{(3)}_{+}(t')$, whereas the  bottom figures of the right columns show the asymptotically vanishing of the expressions  $\int^t dt' \alpha^{(3)}_{-}(t')$. Therefore, according to (\ref{timeint}) our numerical simulations show the asymptotically conservation of the charge $Q^{(3)}_{-}$ and the conservation of the charge $Q^{(3)}_{+}$, within numerical accuracy.  
 
In the Fig. 5 we show  the results for the 
collision of unequal and opposite velocity kinks with parameters $v_2=0.7, v_1=-0.5$ and $\epsilon = 0.03$. Similarly, these simulations of the deformed SG model are performed considering as the initial condition two analytical solitary wave solutions. These results show that the anomalies $\alpha^{(3)}_{\pm}(t)$ do not vanish and the time integrated anomalies $\int^t dt' \alpha^{(3)}_{\pm}(t')$ vanish only asymptotically. According to (\ref{timeint}), the Fig. 5  shows numerically that the both charges  $Q^{(3)}_{\pm}$ are conserved  only asymptotically,  provided that the soliton configuration breaks the even parity associated to the space-reflection symmetry (i.e. it describes a kink-antikink collision with different velocities).

\begin{figure}
\centering
\label{fig4}
\includegraphics[width=2cm,scale=6, angle=0,height=6cm]{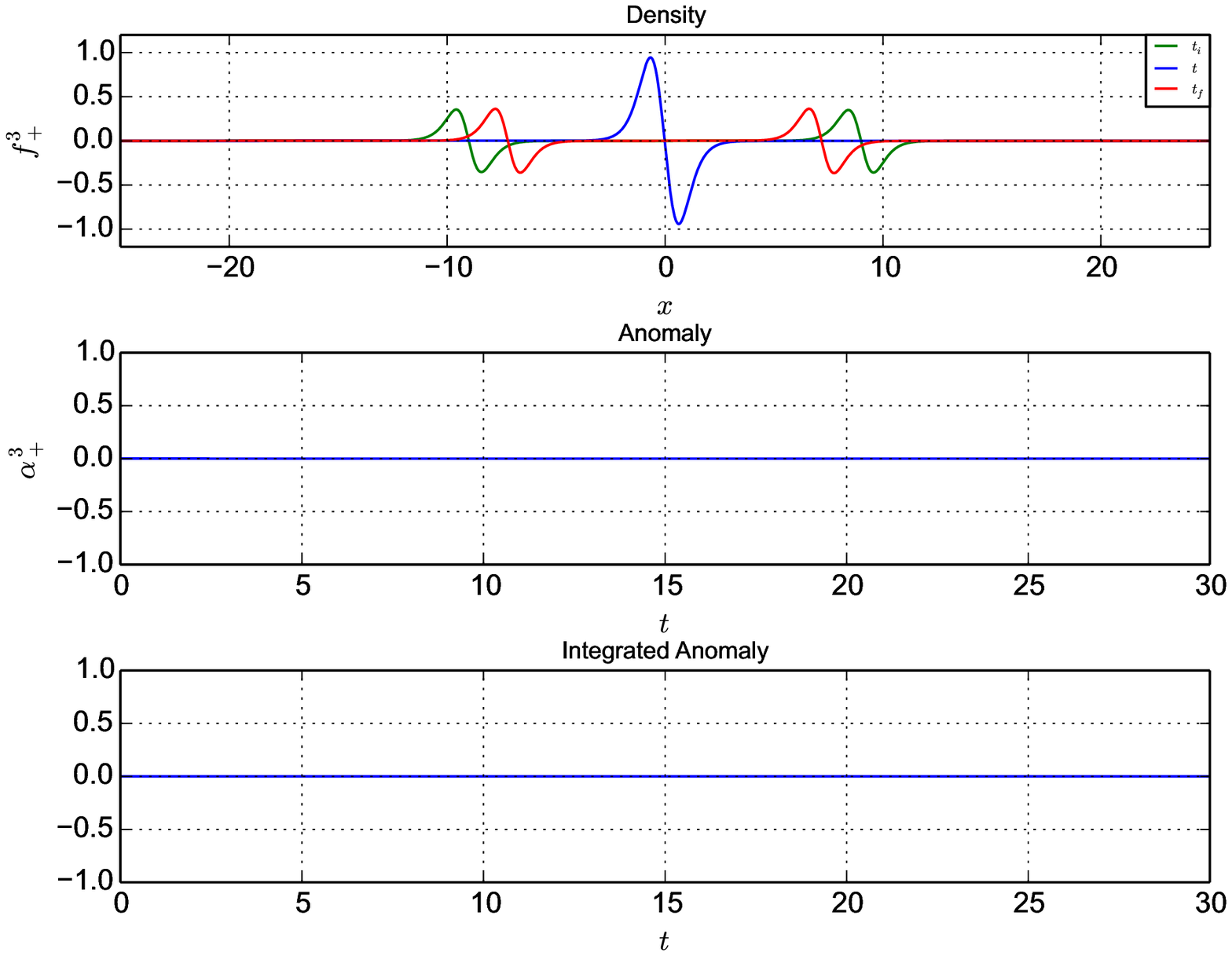}
\includegraphics[width=2cm,scale=6, angle=0,height=6cm]{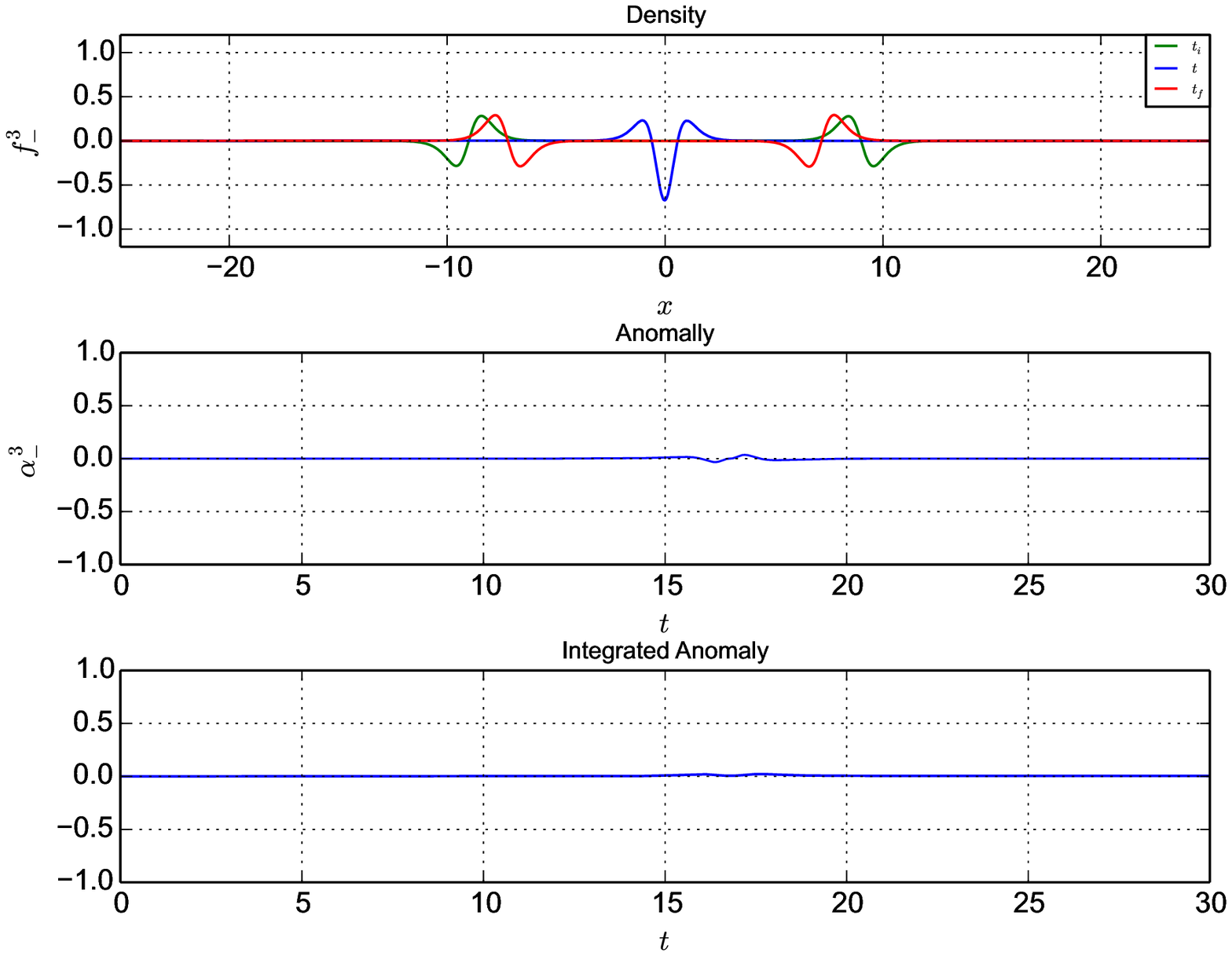}
\parbox{6in}{\caption{(color online)   Left column figures for (\ref{alf11})-(\ref{alf112}), from top to bottom, anomaly density $f^{(3)}_{+}$, anomaly $\alpha_{+}^{(3)}$ and time integrated anomaly  of kink-antikink collision with equal and opposite velocities $v_2=-v_1=0.5$ and $\epsilon = 0.06$. The right column figures show the relevant results for (\ref{alf223})-(\ref{alf2233}) related to anomaly $\alpha_{-}^{(3)}$. The density figures in the both columns correspond to initial (green), collision (blue) and final (red) configurations of the kink-antikink scattering.}}
\end{figure}

\begin{figure}
\centering
\label{fig8}
\includegraphics[width=2cm,scale=6, angle=0,height=6cm]{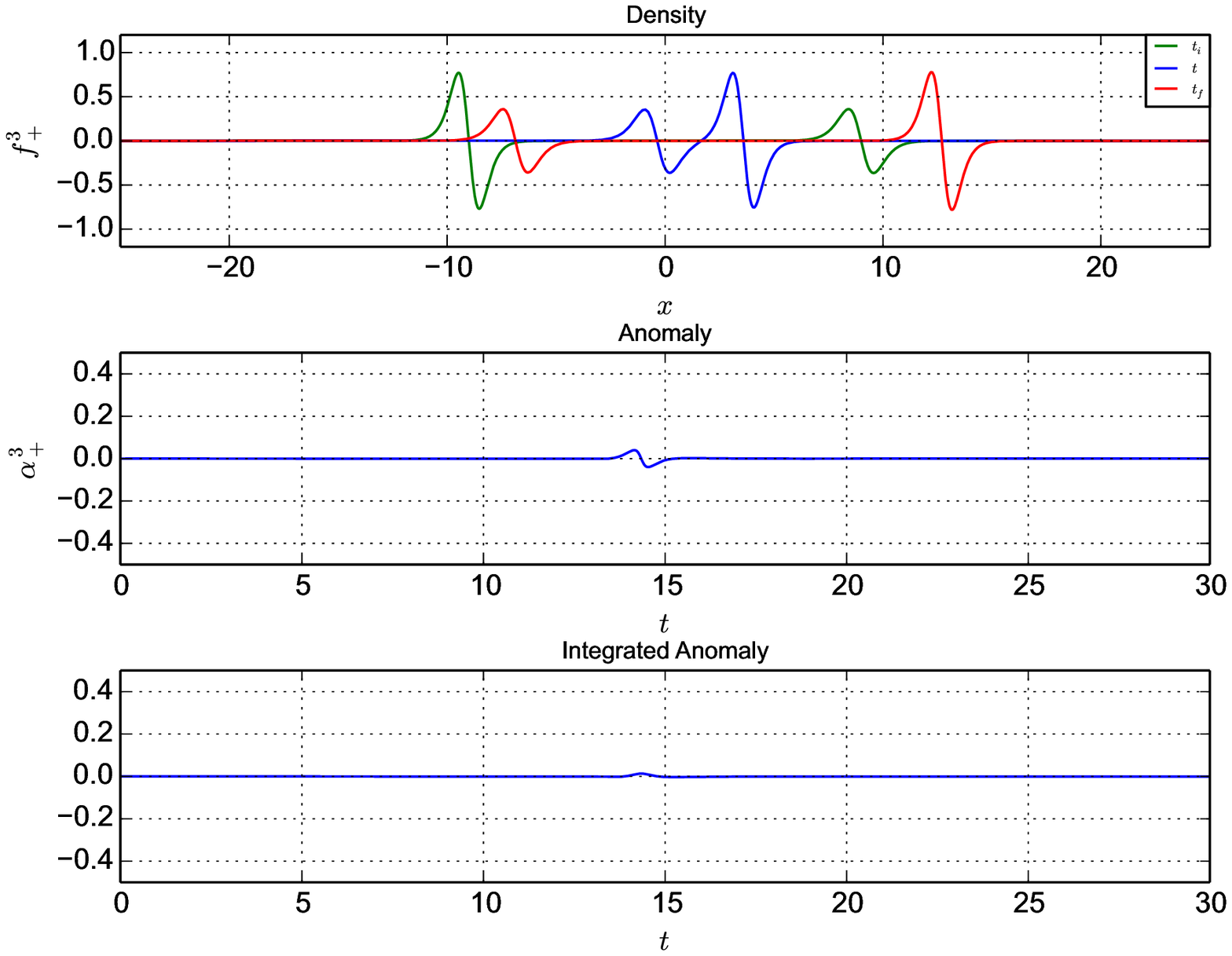}
\includegraphics[width=2cm,scale=6, angle=0,height=6cm]{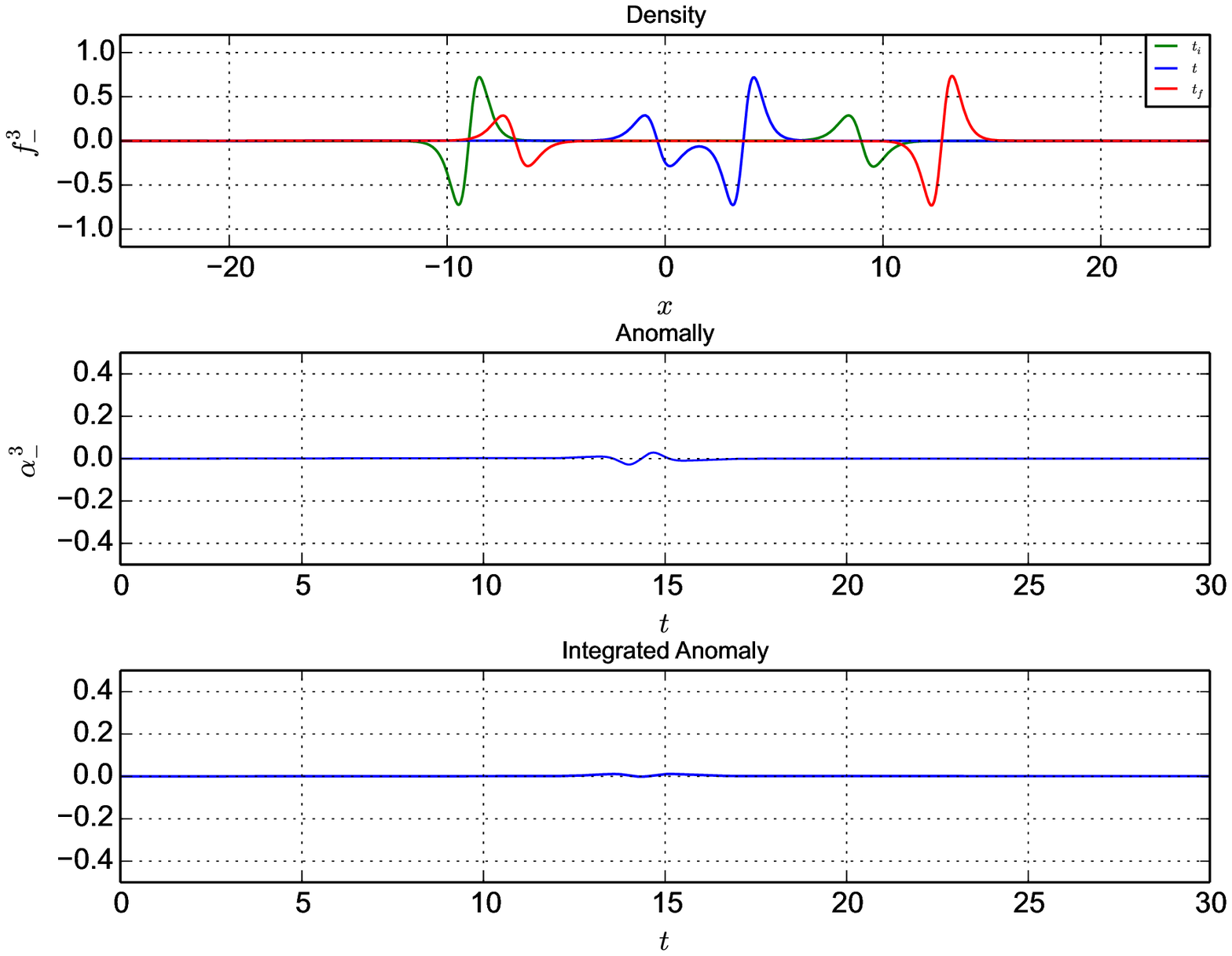}
\parbox{6in}{\caption{(color online)  Left column corresponds to  $\alpha_{+}^{(3)}$ and right column to $\alpha_{-}^{(3)}$ for kink-antikink collision with different velocities $v_2=0.7$ (left soliton),\,$v_1=-0.5$ (right soliton) and $\epsilon = 0.03$.}}
\end{figure}
 
\[
\]

\subsection{kink-kink}

The simulations of the kink-kink system of the deformed SG model will consider, as the initial condition, two analytical solitary wave solutions. In fact, in order to have a  kink-kink system for $t=0$ we consider two kinks ($\eta_1= 1,\eta_2=1, l=0$),  according to the solution in eq. (1.2) of \cite{jhep1}, located some distance apart and stitched together at the middle point $x=0$.  
 
These kink-kink  simulations  are presented in the Figs. 6-7. In the Fig. 6 we show  the results for the 
collision of equal and opposite velocity  solitons with  parameters $v_2=-v_1=0.5$ and $\epsilon = 0.06$. The relevant anomaly densities $f^{(3)}_{\pm}(x,t)$, anomalies $\alpha^{(3)}_{\pm}(t)$, as presented in (\ref{alf11})-(\ref{alf112}) and (\ref{alf223})-(\ref{alf2233}), respectively,  and the time integrated anomalies $\int^t dt' \alpha^{(3)}_{\pm}(t')$ as functions of time, are plotted. The relevant anomaly densities $f^{(3)}_{\pm}(x,t)$ in (\ref{alf112}) and (\ref{alf2233}) have been plotted as functions of $x$ for three successive times (top figures of each column). These plots show qualitatively the behaviour of these functions such that their integration in the whole space furnishes vanishing  $\alpha^{(3)}_{+}(t)$ and non-vanishing  $\alpha^{(3)}_{-}(t)$ anomalies (middle figures in the both columns). The bottom figures of the left columns show the vanishing expressions $\int^t dt' \alpha^{(3)}_{+}(t')$, whereas the  bottom figures of the right columns show the asymptotically vanishing of the expressions  $\int^t dt' \alpha^{(3)}_{-}(t')$. Therefore, according to (\ref{timeint}) our numerical results show the asymptotically conservation of the charge $Q^{(3)}_{-}$ and the conservation of the charge $Q^{(3)}_{+}$, within numerical accuracy. Thus, these results provide qualitatively similar behaviour to the  kink-antikink collision anomalies  simulated in the previous subsection.

In the Fig. 7  we show  the results for the 
collision of unequal and opposite velocity kink-kink collisions with parameters $v_2=0.9, v_1=-0.4$ and $\epsilon = 0.06$. Similarly, these simulations of the deformed SG model are performed considering as the initial condition two analytical solitary wave solutions. These results show that the anomalies $\alpha^{(3)}_{\pm}(t)$ do not vanish and the time integrated anomalies $\int^t dt' \alpha^{(3)}_{\pm}(t')$ vanish only asymptotically. Therefore, in the Fig. 7 we have shown numerically that the  charges $Q^{(3)}_{\pm}$ are only asymptotically conserved when
the configuration does not possess a space-reflection symmetry (kink-kink collision with different velocities).

Therefore, according to the above numerical results,  we may conclude that the even(odd) parity associated to the space-reflection symmetry of the kink-antikink (kink-kink) configuration is a necessary condition in order to have a conserved $Q^{(3)}_{+}$ charge, within the numerical accuracy of our simulations. In fact, the Fig. 5 shows the non-vanishing of the anomaly  $\alpha^{(3)}_{\pm}(t)$ computed for kink-antikink configuration without even space-reflection parity. In addition,  the  Fig. 7 shows the non-vanishing of these anomalies computed for kink-kink configuration without odd space-reflection parity.

\begin{figure}
\centering
\label{fig10}
\includegraphics[width=2cm,scale=6, angle=0,height=6cm]{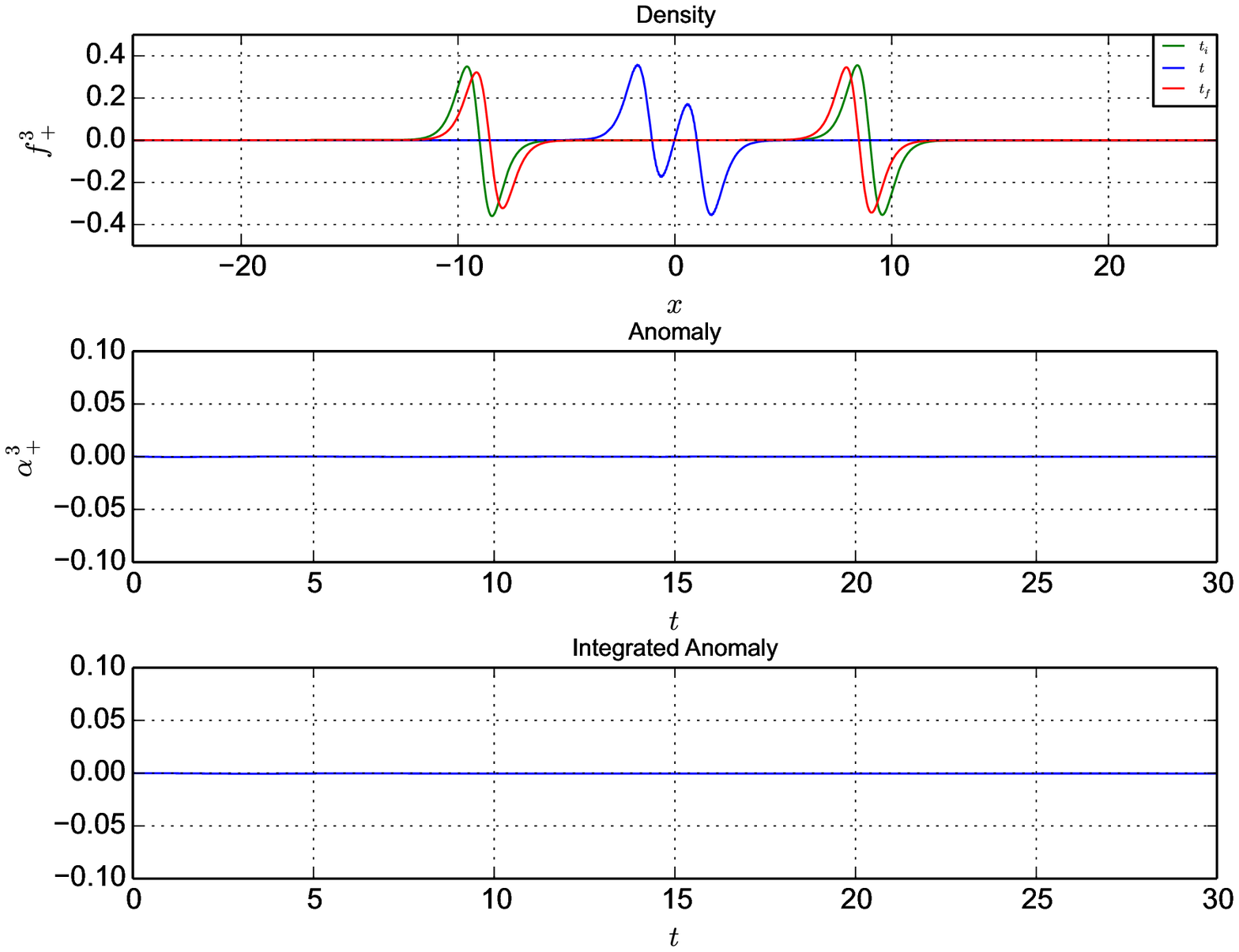}
\includegraphics[width=2cm,scale=6, angle=0,height=6cm]{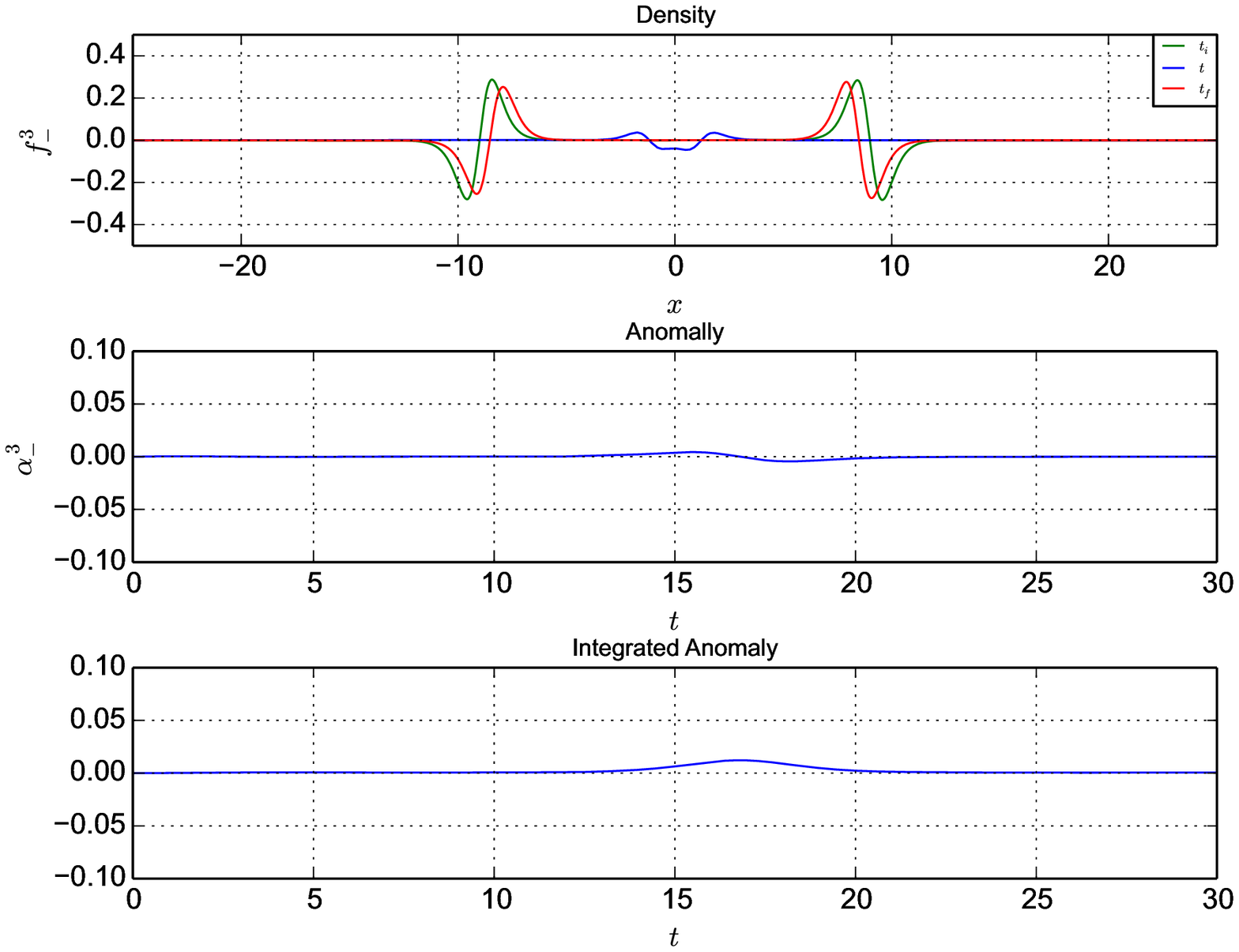}
\parbox{6in}{\caption{(color online) Left column figures for (\ref{alf11})-(\ref{alf112}), from top to bottom, anomaly density $f^{(3)}_{+}$, anomaly $\alpha_{+}^{(3)}$ and time integrated anomaly  of kink-kink collision with equal and opposite velocities $v_2=-v_1=0.5$ and $\epsilon = 0.06$. The right column figures show the relevant results for (\ref{alf223})-(\ref{alf2233}) related to anomaly $\alpha_{-}^{(3)}$. The density figures in the both columns correspond to initial (green), collision (blue) and final (red) configurations of the kink-kink scattering.}}
\end{figure}

\begin{figure}
\centering
\label{fig14}
\includegraphics[width=2cm,scale=6, angle=0,height=6cm]{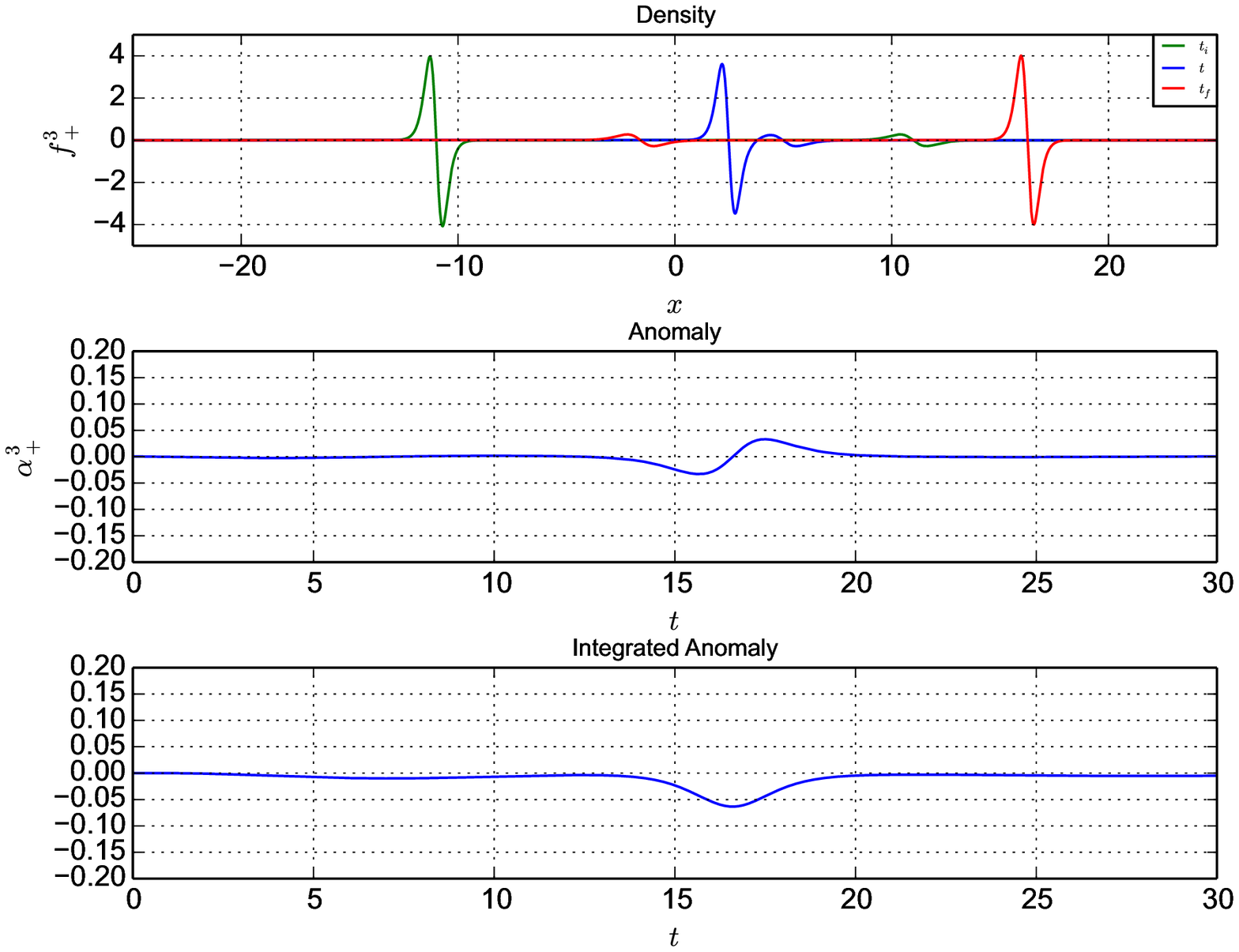}
\includegraphics[width=2cm,scale=6, angle=0,height=6cm]{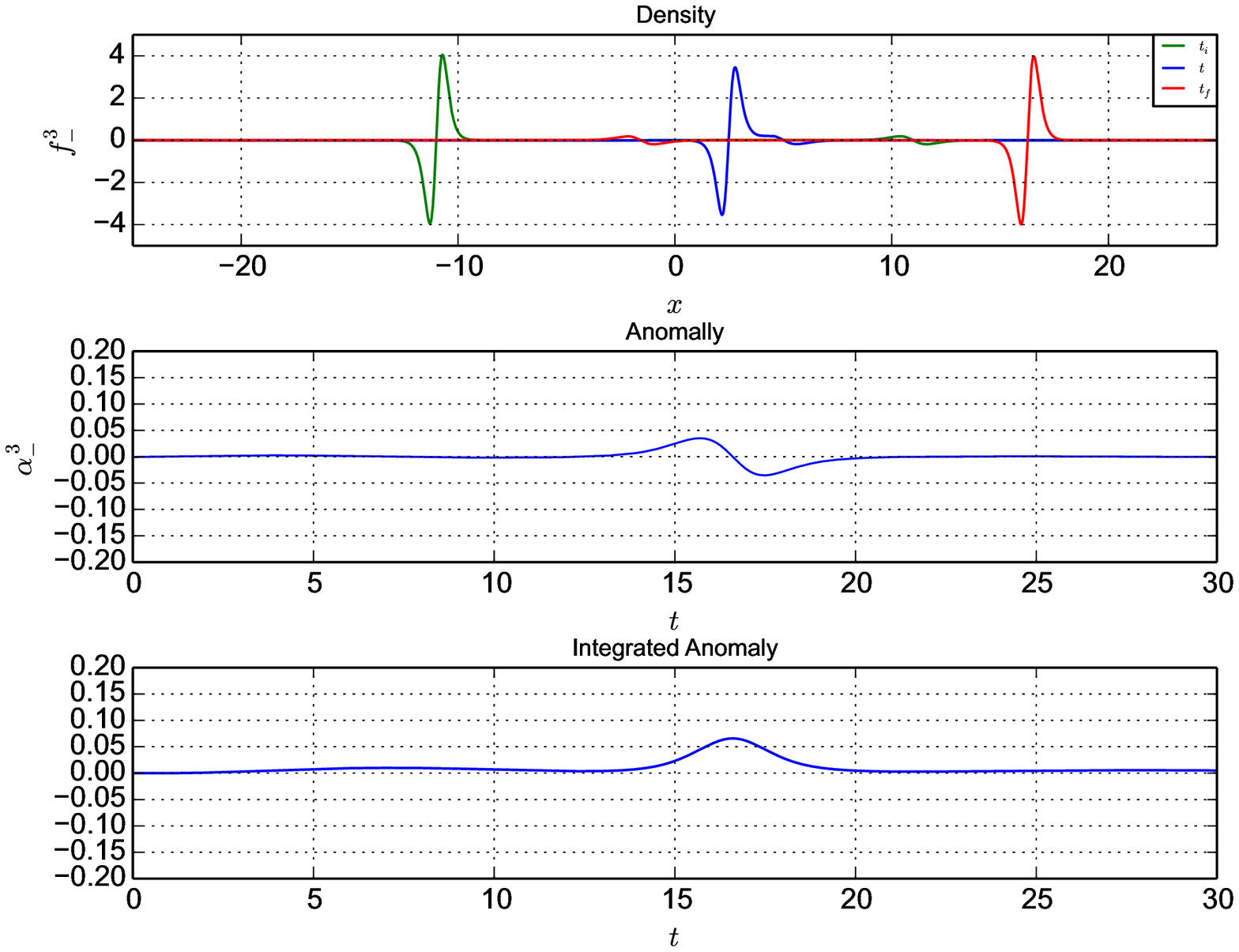}
\parbox{6in}{\caption{(color online) Left column corresponds to  $\alpha_{+}^{(3)}$ and right column to $\alpha_{-}^{(3)}$ for kink-kink collision with different velocities $v_2=0.9$ (left soliton),\,$v_1=-0.4$ (right soliton) and $\epsilon = 0.06$.}}
\end{figure}

However, for the both kink-antikink and kink-kink solitons of the SG model with opposite and different velocities (\ref{2kasym1})-(\ref{2kasym2}) we have shown  that in the center-of-mass reference frame ($x', t'$) provided in (\ref{lor1})-(\ref{lor2}) the parity symmetries are recovered, as discussed in the last paragraph of sec. \ref{spref}. So, the simulations performed in these reference frames, in the both kink-antikink and kink-kink cases, will provide vanishing  $\alpha_{+}^{(3)}$ anomalies as they have been shown above.

\subsection{Breather: kink-antikink bound state}

Some properties of the  breather-like configurations of the deformed sine-Gordon models have been studied in \cite{jhep1,jhep3, arxiv1} through numerical simulations. In \cite{jhep1} the breather-like configurations were obtained by numerically evolving a kink/anti-kink pair initially located some distance apart, as they tend to attract to each other, then  this pair progressively becomes a bound state once the radiation emitted by the configuration is absorbed at the extreme regions of the grid. The second approach, adopted in \cite{jhep3, arxiv1}, considers the analytical sine-Gordon breather solution  and its time derivative evaluated at $t=0$ as the initial condition for the breather simulation of the deformed model. This procedure also produced some radiation which has been absorbed at the edges of the grid. Here we adopt the second approach which seems to be  more appropriate to our purposes; for example, the initial condition will be an analytical function of the SG breather at rest and it already  possesses the positive parity of the breather-like configuration which we are looking for. In addition, the initial energy expression in terms of the frequency turns out to be equal to the one associated to the usual SG model, as we will see below. Nevertheless, the first approach deserves a careful examination in the context of our constructions and we will postpone it for a future work. 

In the Figs. 8-11 the simulations for the breather-like  solution of the deformed SG model are presented. As mentioned above, the input for our program will be the SG breather  solution at rest (\ref{br1}), and so the initial condition for our simulations becomes 
\br
\label{br2}
\vp_0^{b}|_{t=0} = 0,\,\,\,\,\frac{d}{dt}\vp_0^{b}|_{t=0} = \frac{\sqrt{1-\nu^2_0}}{\cosh{(\sqrt{1-\nu^2_0}\, x)}}. 
\er 
For this initial configuration we have $\frac{d}{dx}\vp_0^{b}|_{t=0} =0 $, and $V(\vp=0) = 0 $, and then from (\ref{ener}) one has that the initial energy $E_0 = Q^{(1)}_{+}$ becomes 
\br
\label{eneri}
E_0 =  \sqrt{1-\nu^2_0}.
\er
In fact, this is the initial energy of the sine-Gordon breather. So, our simulations of the deformed SG model consider as the initial condition the analytical breather solution presented in Fig. 1 b.  We will consider the parameters $\nu_0 = 0.8944$  and $\epsilon = \pm 0.06, \pm 0.03$. The initial energy corresponding to this frequency becomes $E_0  = 0.4473 $ and the amplitude of the breather configuration is $A_0 =\vp_0^{b}(0, \frac{\pi}{2 \nu_0}) =   0.46371$. The relevant anomaly densities $f^{(3)}_{\pm}(x,t)$, anomalies $\alpha^{(3)}_{\pm}(t)$, as presented in (\ref{alf11})-(\ref{alf112}) and (\ref{alf223})-(\ref{alf2233}), respectively,  and the time integrated anomalies $\int^t dt' \alpha^{(3)}_{\pm}(t')$ as functions of time, are plotted. The relevant anomaly densities $f^{(3)}_{\pm}(x,t)$ in (\ref{alf112}) and (\ref{alf2233}), respectively,  have been plotted as functions of $x$ for three successive times (top figures of each column). These plots show qualitatively the behaviour of these functions such that their integration in the whole space furnishes vanishing  $\alpha^{(3)}_{+}(t)$ and non-vanishing (periodic in time)  $\alpha^{(3)}_{-}(t)$ anomalies (middle figures in the both columns). The bottom figures of the left columns show the vanishing expressions $\int^t dt' \alpha^{(3)}_{+}(t')$, whereas the  bottom figures of the right columns show that the expressions  $\int^t dt' \alpha^{(3)}_{-}(t')$ are periodic in time. Therefore, according to (\ref{timeint}) our numerical results show the oscillation of the charges $Q^{(3)}_{-}$ around a fixed value and the exact conservation of the charges $Q^{(3)}_{+}$, within numerical accuracy.  
 
A sensible definition of the ``lifetime" of a breather in deformed SG models is not yet available; however, a working definition has been put forward recently in relation to a particular deformation of the SG model ( see \cite{arxiv1} and references therein). They consider ``short-lived" and ``long-lived" breathers according to the energy loss of the field configuration in $4 \times 10^4$ units of time. The  ``short-lived" system was characterized by the energy loss of more than $10 \%$ during this time, and the  ``long-lived" system if it was less than $2 \%$. So, following  this terminology   our simulations for $|\epsilon|=0.03, 0.06$ and $\nu_0 =0.8944 $ can be termed as ``long-lived" breathers, since for the time of the order of $10^5$ one has an energy  loss of less than $2\%$. In fact, from the data in Fig. 9 for $|\epsilon| = 0.03$ one has $\frac{(E_0-E_f)}{E_0} \times 100 \approx 0.05\% $, where $E_f \approx 0.4471$. Similarly, the data presented in Fig. 9 for $|\epsilon|= 0.06$ and $\nu_0 =0.8944 $ show the energy loss of the order of $\sim 0.1\%$, thus showing that in this case we also have ``long-lived" breathers. So, we are confident on our extensive numerical simulations showing the existence of long-lived breathers in the region $|\epsilon| < 0.1$. Moreover, we noticed that for some set of parameters $\epsilon$ and $\nu_0$, and after about $ 5\times 10^4$ units of time, the quasi-breathers start moving slowly, and thus break their even parity symmetry. This phenomenon lies beyond our present scope and deserves future investigations.      

Notice that the time dependence of the energies of the breather systems presented in Fig. 9 resemble qualitatively to the ones studied in \cite{arxiv1}, when a relevant deformation of SG was considered such that the space-time parity symmetry holds (see e.g. their Figs. 7 and 9). The authors in \cite{arxiv1} analysed the time oscillating component $\alpha^{(2n+1)}$ as presented in (\ref{qsc1}) of the present paper, instead  our numerical results consider the both composed anomalies $\alpha_{\pm}^{(3)}$. In fact, our numerical simulations show that the vanishing of the composed anomaly $\alpha_{+}^{(3)}$ leads to the exactly conserved charge $Q^{(3)}_{+}$, whereas the other composed anomaly $\alpha_{-}^{(3)}$ oscillates in time (see Fig. 8), and then the relevant charge $Q^{(3)}_{-}$ will oscillate around a fixed value, according to the equations (\ref{lcom}) and (\ref{timeint}).           
 
\begin{figure}
\centering
\label{fig18}
\includegraphics[width=2cm,scale=6, angle=0,height=6cm]{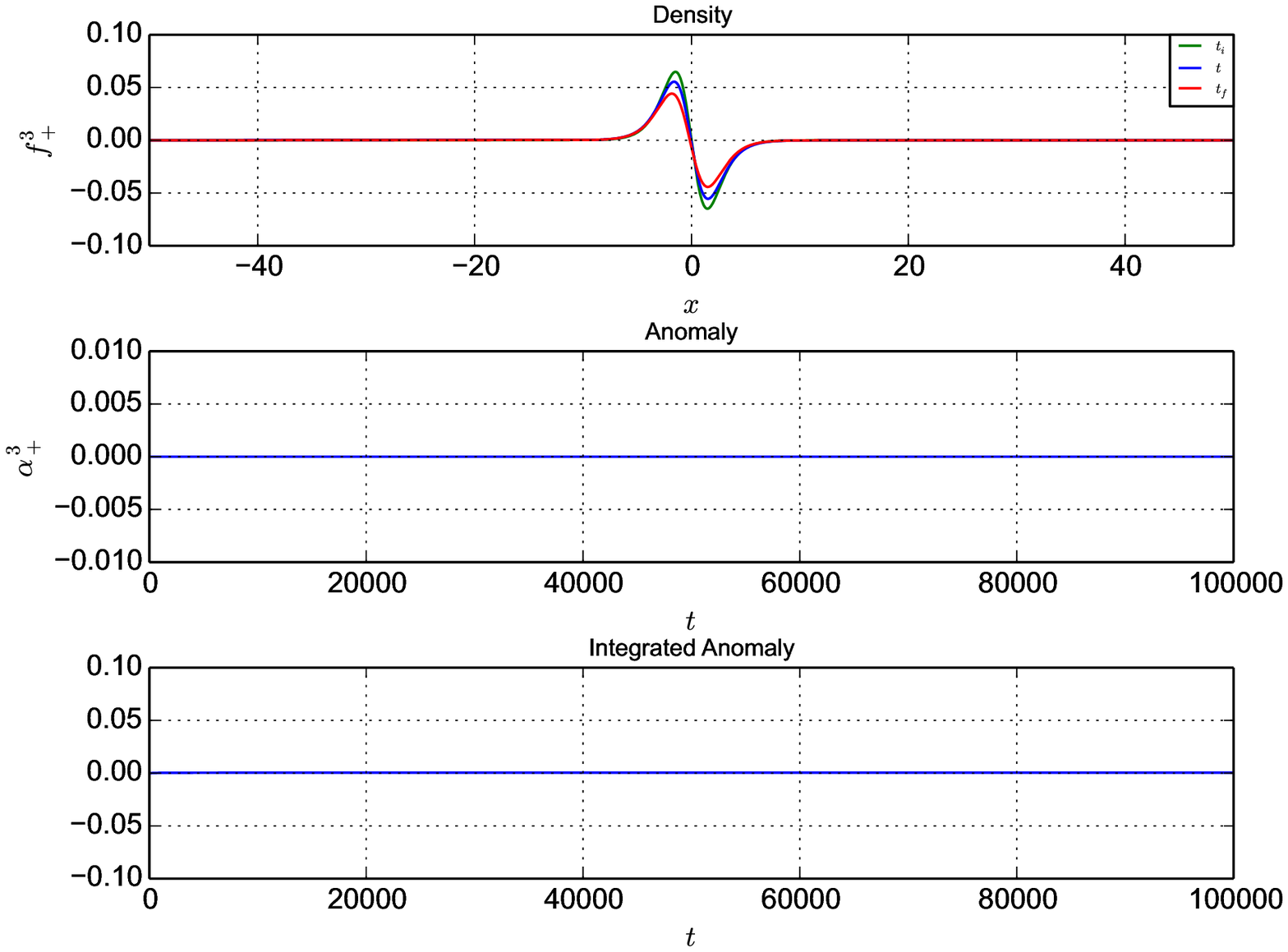}
\includegraphics[width=2cm,scale=6, angle=0,height=6cm]{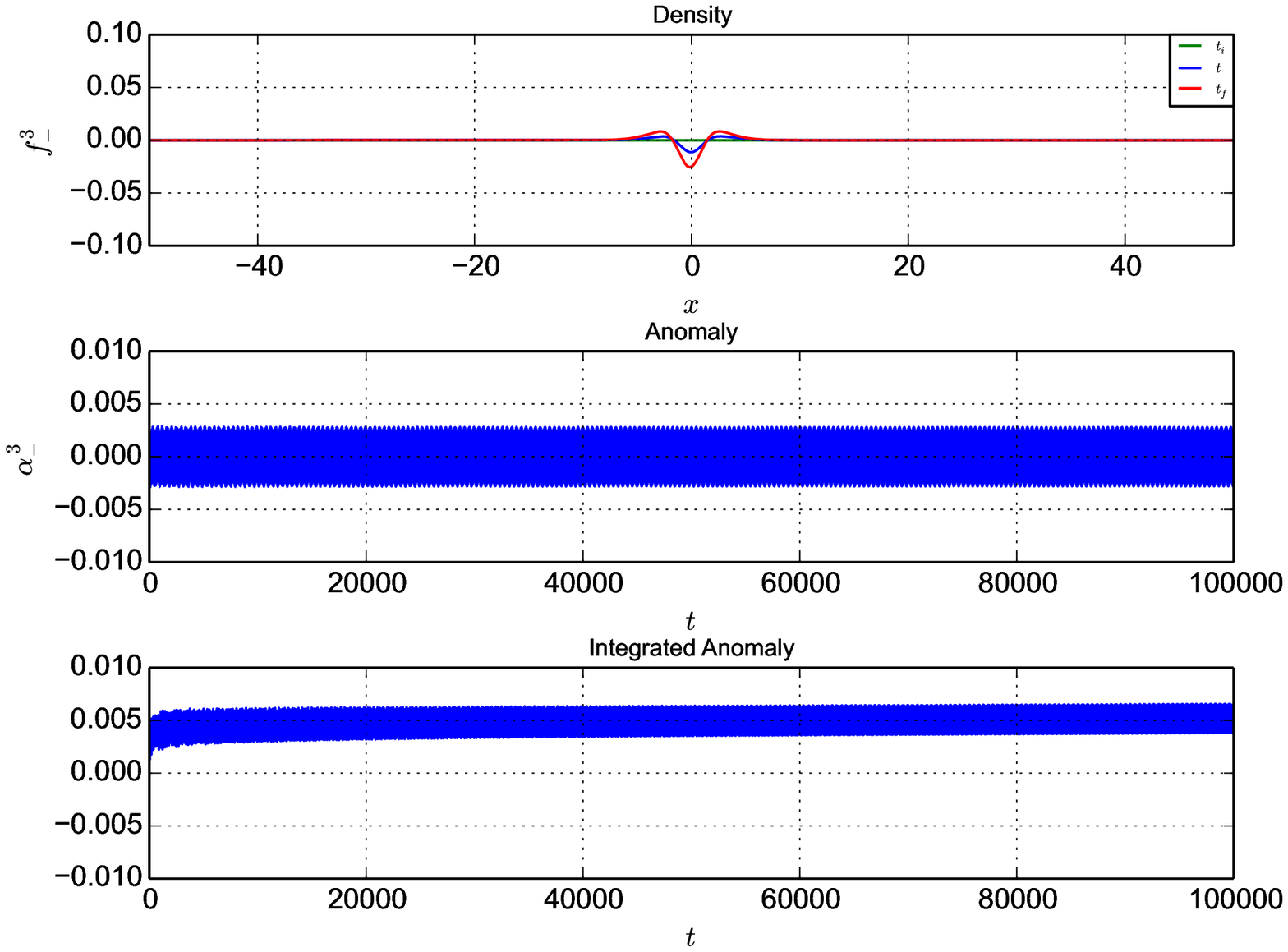}
\parbox{6in}{\caption{(color online)  Left column figures for (\ref{alf11})-(\ref{alf112}), from top to bottom, anomaly density $f^{(3)}_{+}$, anomaly $\alpha_{+}^{(3)}$ and time integrated anomaly   for the breather (kink-antikink bound state) oscillation with $\epsilon = 0.06$ and initial condition frequency $\nu_0=0.8944$ in (\ref{br2}). The right column figures show the relevant results for (\ref{alf223})-(\ref{alf2233}) related to anomaly $\alpha_{-}^{(3)}$. The density figures in the both columns correspond to three successive times ($t_i < t< t_f$) in the interval $[t_f -T_0, t_f]$,\, with  $T_0=7.025$. The long-lived breather is characterized by the final time of the order $t_f \approx 10^5$.}}
\end{figure}

\begin{figure}
\centering
\label{fig20}
\includegraphics[width=2cm,scale=6, angle=0,height=6cm]{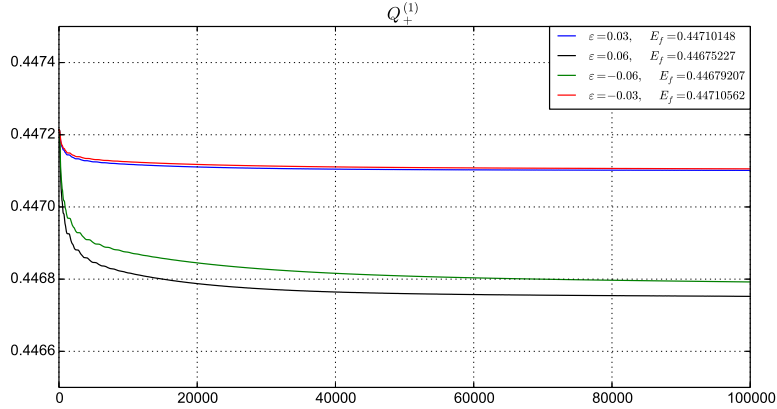} 
\parbox{6in}{\caption{(color online)   Breather simulations of the deformed SG theory (\ref{eq1})-(\ref{dpot}) with  initial configuration (\ref{br2}), deformation and initial condition parameters given by  $\epsilon = \pm 0.06,\, \pm 0.03$ and  $\nu_0=0.8944,\,T_0=7.025$, respectively. The plots show the time dependence of the energy $E=Q_{+}^{(1)}$ (\ref{ener}) (in adimensional units). The final energies $E_f$ of the stabilized breathers are lower than the initial energy $E_0 = 0.4473$ of the SG breather. Notice that the final  time of the order of $t_f \approx 10^5$ characterizes the ``long-lived" breathers.}}
\end{figure}

\begin{figure}
\centering
\label{fig21}
\includegraphics[width=2cm,scale=6, angle=0,height=6cm]{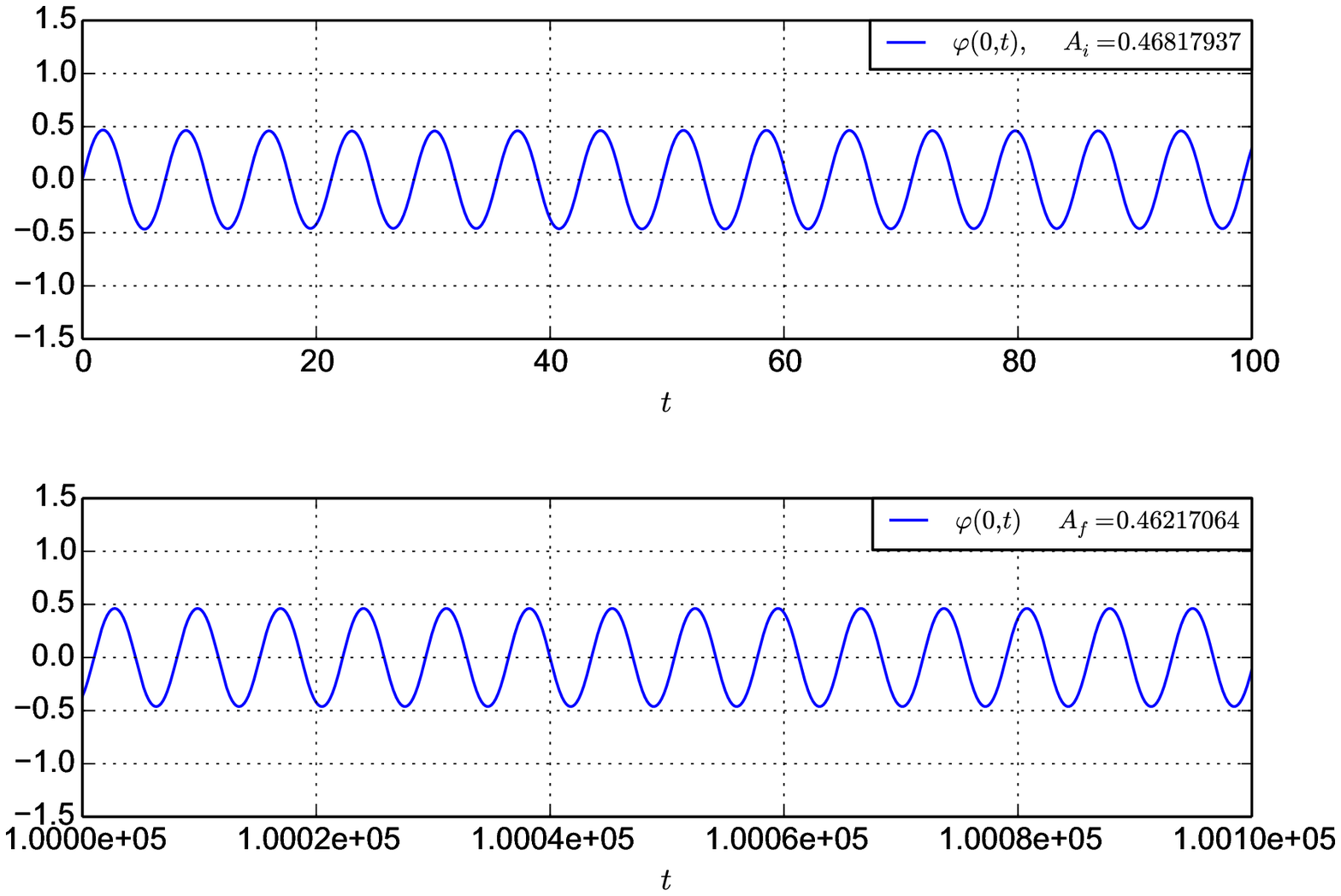} 
\includegraphics[width=2cm,scale=6, angle=0,height=6cm]{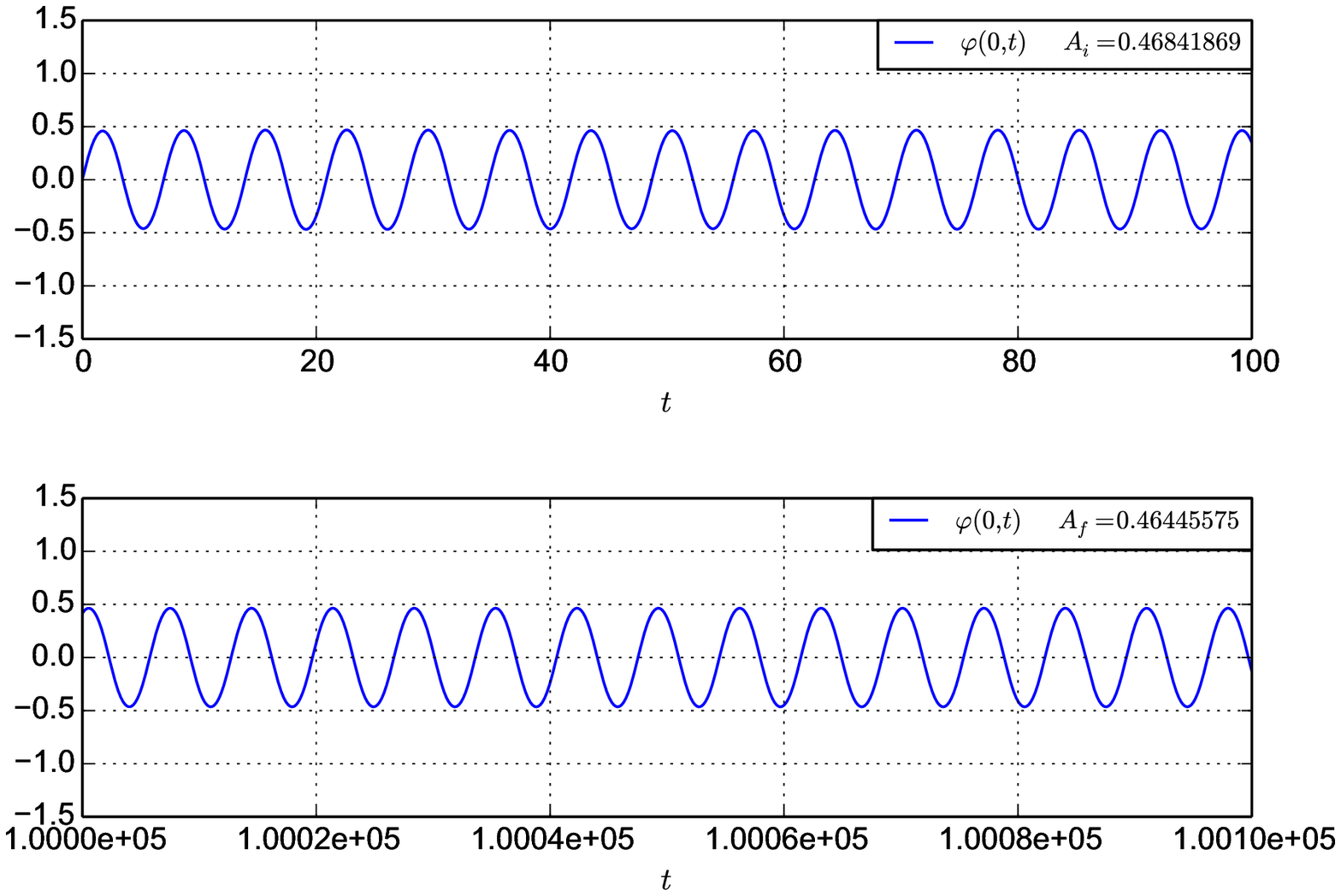} 
\parbox{6in}{\caption{(color online)  Breather oscillations at $x=0$ during the initial $i=[0, 10^2]$ (top figures) and final  $f=[10^5, 1.0010 \times 10^5]$ (bottom figures) intervals of time with parameters $\epsilon = 0.03$ (left column) and $\epsilon = -0.03$ (right column), respectively. The frequency increased $(\nu_i < \nu_f)$ and the amplitude decreased ( $A_i > A_f $ ) in each case. The $\nu_i\,'s$ are averaged over several periods of time and the $\nu_f\,'s$ are the final stabilized frequencies.}}
\end{figure}

\begin{figure}
\centering
\label{fig22}
\includegraphics[width=2cm,scale=6, angle=0,height=6cm]{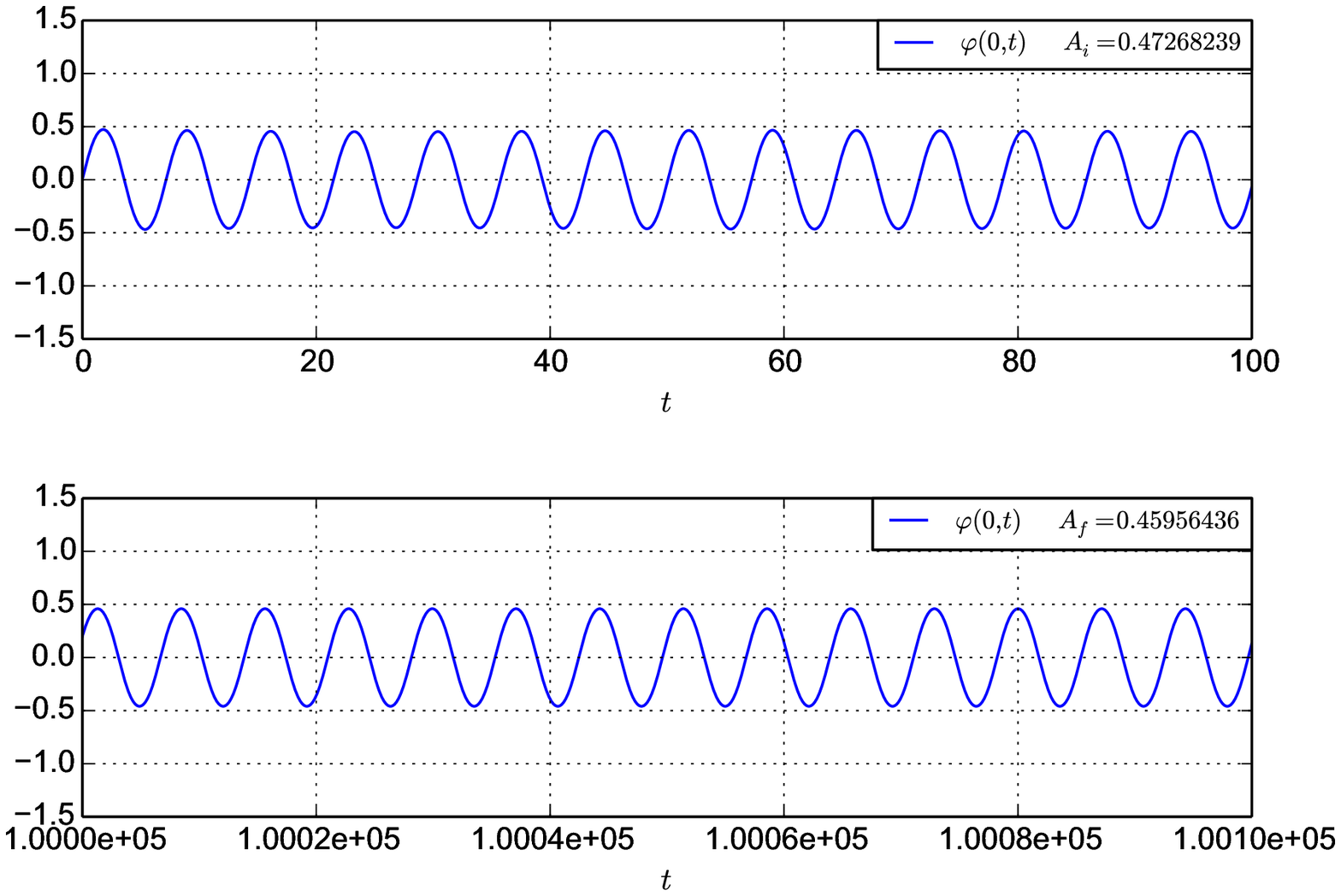} 
\includegraphics[width=2cm,scale=6, angle=0,height=6cm]{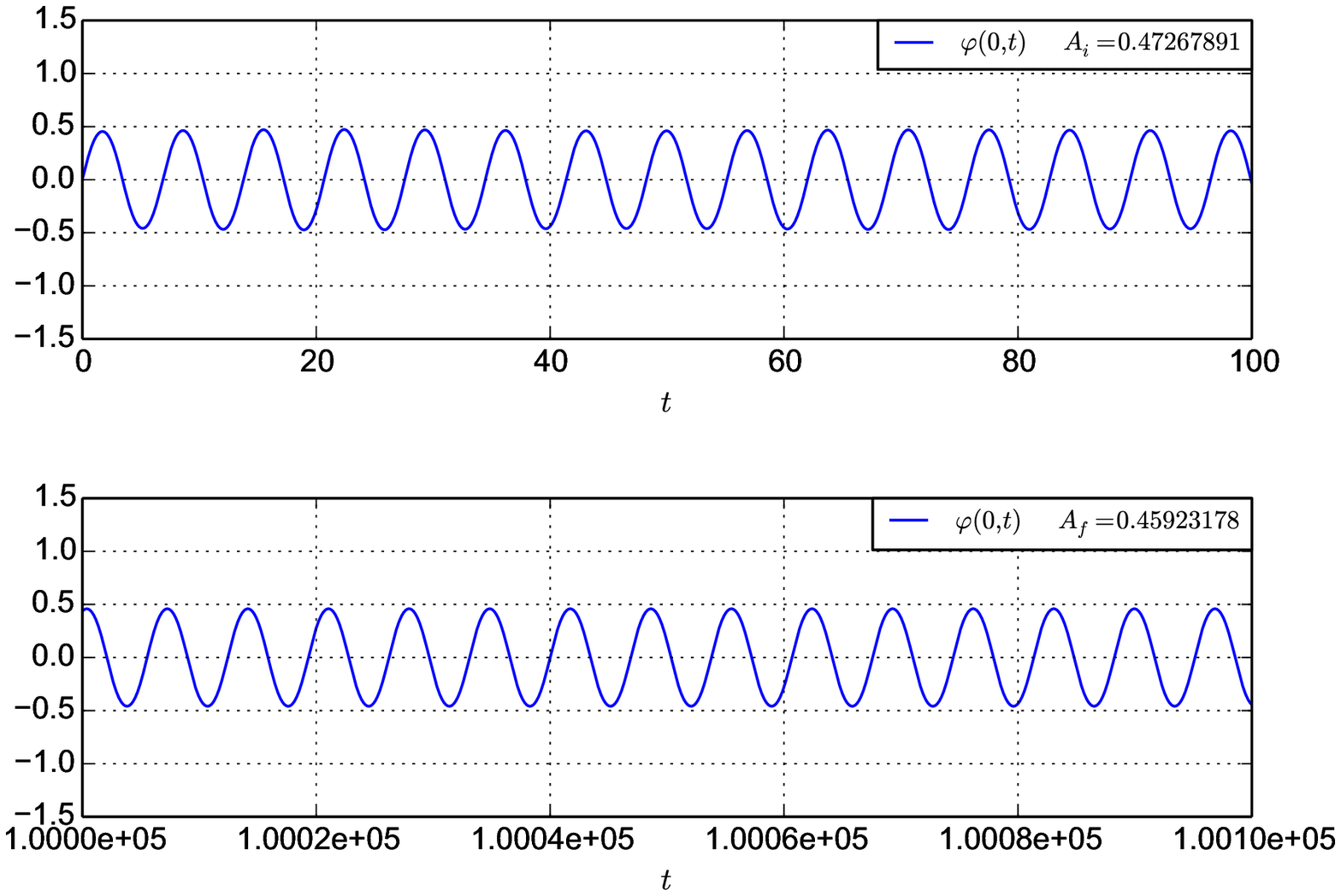} 
\parbox{6in}{\caption{(color online)  Breather  oscillations at $x=0$ during the initial $i=[0, 10^2]$ (top figures) and final  $f=[10^5, 1.0010 \times 10^5]$ (bottom figures) intervals of time with parameters $\epsilon = 0.06$ (left column) and $\epsilon = -0.06$ (right column), respectively. The frequency increased $(\nu_i < \nu_f)$ and the amplitude decreased ( $A_i > A_f $ ) in each case.  The $\nu_i\,'s$ are averaged over several periods of time and the $\nu_f\,'s$ are the final stabilized frequencies.}}
\end{figure}

Finally, let us see what would happen if the energy dependence of the stabilized breather on the frequency is assumed to be similar to the usual SG breather as in eq. (\ref{eneri}). So, for the stabilized breather we would have $E_{f} \sim \sqrt{1-\nu^2_{f}}$. Next, consider the relationship
\br
\label{rat}
\(\frac{E_{0}}{E_f}\)^2 = \frac{1-\nu_{0}^2}{1-\nu_{f}^2},
\er
and compute the validity of this ratio for the various simulations. In fact, in Fig. 10  one notices that the breather simulations for $\epsilon=0.03,\,E_0 = 0.4473$ show that the final frequency  becomes $\nu_{f}=0.88617$ and the initial energy decreases to $E_f=0.4471$. For these numerical values the l.h.s. of relationship (\ref{rat}) becomes $\sim 1.0009$, whereas the r.h.s.  $\sim 1.0736$, thus showing that the ratio is a good approximation to within $7\%$. Similar computations, using the data from Fig. 11 for the case $\epsilon = 0.06$, show the validity of (\ref{rat}) to within  $6\%$. Whereas, the l.h.s and r.h.s of (\ref{rat}) computed for the relevant values associated to each of the negative parameters $\epsilon= -0.03$ and $-0.06$ differ to within  $15\%$ and $25\%$, respectively. These results, and some other ones, such as the existence of moving quasi-breathers briefly mentioned above, show that the problems of stability, energy dependence on frequency $\nu$ and the deformation parameter $\epsilon$ for breathers deserve careful analytical and numerical studies. We hope to consider such systems in more detail in the near future.

\section{Discussions and some conclusions}
\label{conclu}

The well known results of \cite{jhep1} for soliton solutions can be summarized as follow: a) the model  possesses an infinite number of exactly conserved quantities for one-soliton type solutions, i.e. for kink-type solitary waves traveling with a constant speed. b) For two-soliton (kink-kink, kink-antikink) type solutions possessing a special space-time parity symmetry (\ref{stsym1})-(\ref{stsym2}), the charges are asymptotically conserved. This means that these quantities  vary in time during the collision process (and sometimes can vary quite a lot) of two one-solitons but return, in the distant future (after the collision), to the values they had in the distant past (before the collision). c) For breather-type solutions, with the same space-time parity symmetry, those charges oscillate around a fixed value. 

We have shown that the quasi-conserved charges of the deformed sine-Gordon model (\ref{eq1}) studied in \cite{jhep1} split into two subsets, with different conservation properties. Through linear combinations, at each level $n$ in (\ref{charlc}), of the dual set of asymptotically conserved charges we have obtained a subset comprising a new infinite tower of exactly conserved charges (\ref{cons1}), and a second subset containing the remaining asymptotically conserved ones. The two-soliton (kink-kink and kink-antikink) and breather-like (kink-antikink bound state)  solutions in their center-of-mass frame  possessing a special space-reflection parity symmetry (\ref{px})-(\ref{pxvp}) give rise to the new tower of exactly conserved charges. In order to examine the parity symmetry and the vanishing of the anomalies (\ref{ano123}) we have used a technique involving the space-reflection and an order two $\IZ_2$ automorphism $\widetilde{\Sigma}$  (\ref{aut22}) of the  $sl(2)$ loop algebra, in the context of dual anomalous zero-curvature (Lax) representations of the deformed model.

In sec. \ref{sec:expansion} we have implemented a perturbation theory, by a power series expansion on $\epsilon$, in order to study the interplay between the space-reflection symmetry and dynamics of the solutions.  In this context, the zeroth order solution in  $\epsilon$ can be chosen to be the two-soliton solutions (kink-antikink, kink-kink and breather) of the integrable sine-Gordon  model, which 
satisfy the symmetries (\ref{pxxy})-(\ref{pxx1}). We have provided an order by order prescription to construct a solution of the deformed model satisfying the parity property, i.e. even or odd parity solution, which implies the existence of a tower of exactly conserved charges associated to a field configuration with definite parity. 
    
The vanishing of the anomalies $\alpha^{(2 n+1)}_{\pm}$ (\ref{anolc0}) for a general solitary wave solution has been verified in section \ref{lorentz} by showing that  their components $\alpha^{(2n+1)} dt $ and $\widetilde{\alpha}^{(-2n-1)} dt$  transform as tensors under the Lorentz group. Since for static solitary wave solutions the anomalies vanish, and the moving waves can be obtained through a Lorentz boost, then we may conclude that the anomalies  must vanish in all Lorentz frames.

We have checked the predictions of our analytical calculations through numerical simulations. For the special deformed potential (\ref{dpot})  we have computed the first non-trivial anomalies $\alpha^{(3)}_{\pm}$ of the quasi-conservation laws involving the charges  $Q^{(3)}_{\pm}$. We have verified that the  anomaly $\alpha^{(3)}_{+}$ vanishes, and consequently the exact conservation of the charge $Q^{(3)}_{+}$ holds for various two-soliton configurations, within numerical accuracy. Moreover, the  anomaly $\alpha^{(3)}_{-}$  does not vanish, and then the  charge $Q^{(3)}_{-}$ is asymptotically conserved for the various two-soliton configurations.  
We have verified   numerically  for the collision of kink-antikink and  kink-kink systems traveling in opposite directions with equal velocities, as shown in the Figs. 4-5 and Figs. 6-7, respectively. Moreover, for the breather-type solutions with even parity under space-reflection symmetry (breather at rest) we have constructed a tower of exactly conserved charges and a subset of charges which oscillate around a fixed value (see the Fig. 8). In addition, our extensive numerical simulations show the existence of long-lived breathers in the region $| \epsilon | < 0.1$ (the time dependence of the energy is shown  in Fig. 9 for various $\epsilon$ parameters).  

For  those two-soliton  (kink-antikink and kink-kink) solutions in laboratory coordinates and without the space-reflection parity symmetry we have checked, through numerical simulations, that the both set of charges are only asymptotically conserved (see Fig.5 and Fig.7, respectively). However, in the case of solutions of the usual sine-Gordon model  we have shown that in their center-of-mass reference frames (\ref{lor1})-(\ref{lor2}), the both systems recover their space-reflection symmetries (in this frame they would be suitable solutions in perturbation theory as the zeroth  order in powers of $\epsilon$), and thus the existence of the tower of exactly conserved charges would be guaranteed by the parity symmetry for each pair of solitons in that frame, as shown in the Fig. 4 and Fig. 6, respectively.

The mechanism responsible for the exact conservation of the charges is not well understood yet. As in all of the examples in relativistic field theories where the asymptotically conserved charges have been observed so far, associated to a space-time parity symmetry  \cite{jhep3, jhep4, jhep6}, the two-soliton type (kink-kink, kink-antikink and breather) solutions associated to a sequence of exactly conserved charges present special properties under a space-reflection parity transformation. The only explanation we have found, so far, for this sector of the charges, is that those special soliton-like solutions are eigenstates  under a space-reflection parity transformation for a fixed time, where the point in $x-$coordinate around which space is reversed depends upon the parameters of the particular solution under consideration (in particular, we have set $x_{\Delta}=0$). In the absence of these symmetries in the kink-kink and kink-antikink collisions we have checked numerically the existence of the asymptotically conserved charges only, which allow us to argue that the parity symmetries of the systems of two-solitons in their center-of-mass reference frame is a necessary condition in order to get the set of exactly conserved charges.

Further research work is necessary to settle such
questions as the decay of the breathers and the parameters $\epsilon$ and $\nu$ dependence of the energy of long-lived breathers, which involve the non-linear dynamics of the scattering. The symmetries involved in the quasi-integrability phenomenon deserve further investigation; in particular, the transition between the both phases: broken (laboratory coordinate frame) and unbroken (center-of-mass frame) parity symmetric phases  of the two-soliton solutions and their bound state. In particular, the even parity breather-type system (at rest) for $\epsilon \approx 0.1 $ start moving slowly after $\sim 5\times 10^4$ units of time, and so the parity symmetry is broken and its effect on the tower of exactly conserved charges deserves a careful examination. Finally, the space-time and internal symmetries involved in the quasi-integrability phenomenon deserve further investigation, since  they have potential applications in many areas of non-linear sciences.

\noindent {\bf Acknowledgements:} 
HB thanks Prof. L.A. Ferreira for enlightening discussions on quasi-integrability and FAPEMAT for partial financial support in the initial stage of the work.

\appendix

\section{The expansions}
\label{expansion}
\setcounter{equation}{0}

The expansion of the derivative of the potential becomes
\br
\frac{\partial\,V}{\partial\,\vp} &=& 
\frac{\partial\,V}{\partial\,\vp} \mid_{\ve=0} +
\left[\frac{d\,}{d\,\ve}\(\frac{\partial\,V}{\partial\,\vp}
  \)\right]_{\ve=0}\, \ve + \frac{1}{2} \left[\frac{d^2\,}{d\,\ve^2}\(\frac{\partial\,V}{\partial\,\vp}
  \)\right]_{\ve=0}\, \ve^2 + \ldots\nonumber\\
&=& 
\frac{\partial\,V}{\partial\,\vp} \mid_{\ve=0} +
 \left[\frac{\partial^2 V}{\partial \ve\partial\vp}+
\frac{\partial^2 V}{\partial \vp^2}\,\frac{\partial \vp}{\partial
  \ve}\right]_{\ve=0} \, \ve \\ 
&& + \frac{1}{2} \left[\frac{\partial^3 V}{\partial \ve^2\partial\vp}+ 2
\frac{\partial^3 V}{\pa \ve \partial \vp^2}\,\frac{\partial \vp}{\partial
  \ve} + \frac{\partial^2 V}{\partial \vp^2}\,\frac{\partial^2 \vp}{\partial
  \ve^2}+\frac{\partial^3 V}{\partial \vp^3}\,(\frac{\partial \vp}{\partial
  \ve})^2\right]_{\ve=0}\, \ve^2 + \ldots\nonumber
\er
 
The first few $f_n$'s become
\br
\label{f1}
f_1(\vp_0) &=& -\frac{\partial^2 V}{\partial \ve\partial\vp}|_{\ve=0} ,\\
     &=& \frac{1}{2} \Big[\cos^2{(\vp_0)} + 4 \log{(|\sin{\vp_0}|)} \sin^2{\vp_0}\Big] \sin{(2 \vp_0)},\label{f11}\\
     \label{f2}
 f_2 (\vp_0, \vp_1)&=&  -\frac{1}{2} \left[\frac{\partial^3 V}{\partial \ve^2\partial\vp}|_{\ve=0} + 2
\frac{\partial^3 V}{\pa \ve \partial \vp^2}|_{\ve=0}\, \vp_1 +\frac{\partial^3 V}{\partial \vp^3}|_{\ve=0}\,  \vp_1^2\right] \\
\nonumber
&=&-\frac{1}{2}\Big\{ -4 \sin{(4 \vp_0)}\vp_1^2 - 2 [ \cos^2{ (\vp_0)} + 2 (\cos{(2 \vp_0)}-\cos{(4 \vp_0)}) \log{|\sin{\vp_0}|}] \vp_1\\
\nonumber
&& + \frac{1}{32} \tan{\vp_0} \sec^2{\vp_0} [16+ 23 \cos{(2 \vp_0)} + 8 \cos{(4 \vp_0)} + \cos{(6 \vp_0)} - 16 \log^2{|\sin{\vp_0}|}\times \\
&& ( 3 \cos{(2 \vp_0)} + \cos{(4 \vp_0)})   + 16 (2+ \cos{(2 \vp_0)}) \log{|\sin{\vp_0}|} \sin^2{(2 \vp_0)}]\Big\}\label{f22}.
\er

\br
 f_3 (\vp_0, \vp_1, \vp_2) &=&  -\frac{1}{3 !} \Big[ \frac{\partial^4 V}{\partial \ve^3\partial\vp}|_{\ve=0} + 3
\frac{\partial^4 V}{\pa \ve^2 \partial \vp^2}|_{\ve=0}\, \vp_1 + 6 \frac{\partial^3 V}{\partial \vp^2 \partial \ve}|_{\ve=0}\,  \vp_2 +\nonumber \\
&& \label{f3} 3 \frac{\partial^4 V}{\partial \ve\partial \vp^3}|_{\ve=0}  \vp_1^2 + 6
\frac{\partial^3 V}{\partial \vp^3}|_{\ve=0}\, \vp_1 \vp_2+ \frac{\partial^4 V}{\partial \vp^4}|_{\ve=0}\,  \vp_1^3 \Big] \\
&=& \nonumber -\frac{1}{3!} \Big\{ \frac{1}{16} [-18\sin{(2 \vp_0)}-3\sin{(4 \vp_0)} - 48 (3+ \cos{(2 \vp_0)} ) \sin^2{\vp_0} \\
&& \nonumber \tan{\vp_0} \log{|\sin{\vp_0}|} + 24 (5 \cos{(2 \vp_0)} + \cos{(4 \vp_0)} ) \tan^3{\vp_0} \log^2{|\sin{\vp_0}|} - \\
&& \nonumber 16 (-3 + 5 \cos{(2 \vp_0)} + 2 \cos{(4 \vp_0)} ) \tan^3{\vp_0} \log^3{|\sin{\vp_0}|} ] + \\
&&\nonumber \frac{3}{2} [ 3 - 8 \log{|\sin{\vp_0}|} + 2 \log{|\sin{\vp_0}|} ( -(2+ 6 \log{|\sin{\vp_0}|})\\
&& \nonumber \cos{(2 \vp_0)} + 6 \sec^2{\vp_0}+ 2 \log{|\sin{\vp_0}|}(2 \cos{(4 \vp_0)} -2 \sec^2{\vp_0} + 3 \sec^4{\vp_0}) ) ] \times \\
&& \nonumber \vp_1 +  12[ (\cos{(4 \vp_0)} - \cos{(2 \vp_0)}) \log{|\sin{\vp_0}|}- \cos^2{\vp_0} ] \, \vp_2 - 3 [ 1-\\
&& \nonumber 4 \log{|\sin{\vp_0}|} + \cos{(2 \vp_0)} (4 + 16 \log{|\sin{\vp_0}|} ) ] \sin{(2 \vp_0)} \, \vp_1^2 -\\
&&  24 \sin{(4 \vp_0)}\,  \vp_1 \,\vp_2 - 16 \cos{(4 \vp_0)}\,\vp_1^3  \Big\} .\label{f33}
\er

\end{document}